\documentclass[aps,prb,showpacs,amsmath,amssymb,doublecolumn]{revtex4}
\usepackage{pstricks}
\usepackage{amsmath}
\usepackage{amssymb}
\usepackage{psfrag}
\usepackage{color}
\usepackage{dsfont}
\usepackage{graphicx}
\usepackage{natbib}
\usepackage{helvet}
\DeclareFixedFont{\tr}{OT1}{ptm}{n}{n}{14}
\DeclareFixedFont{\trs}{OT1}{ptm}{n}{n}{10}
\DeclareFixedFont{\trss}{OT1}{ptm}{n}{n}{8}

\newcommand{\av}[1]{\langle #1\rangle}
\newcommand{\nnb}{\nonumber \\}
\newcommand{\taut}{\tilde{\tau}}

\newcommand{\dn}[1]{|#1\rangle}
\newcommand{\ldn}[1]{\langle #1|}
\newcommand{\mat}[2]{|#1\rangle\langle#2|}

\begin{document}

\title{Quantum dynamics in electron-nuclei coupled spin system in
    quantum dots: Bunching, revival, and quantum correlation in
    electron-spin measurements}
\author{ \"Ozg\"ur \c{C}ak{\i}r and Toshihide Takagahara}

\affiliation{Department of Electronics and Information Science,
Kyoto Institute of Technology,
Matsugasaki, Kyoto 606-8585 JAPAN}

\affiliation{ \\}

\affiliation{CREST, Japan Science and Technology Agency, 4-1-8 Honcho,
Kawaguchi,
Saitama 332-0012, JAPAN}

\date{\today}

\begin{abstract}
We investigate quantum dynamics in the electron-nuclei coupled spin system in quantum dots and clarify the
fundamental features of quantum correlation induced via successive electron spin measurements. This quantum
correlation leads to interesting phenomena such as the bunching of outcomes in the electron spin measurements
and the revival of an arbitrary initial electron spin state. The nuclear spin system is also affected by the
quantum correlation and is in fact squeezed via conditional measurements or postselection. This squeezing
is confirmed by calculating the increase in the purity of the nuclear spin system. Thus the successive electron spin
measurements provide a probabilistic method to squeeze the nuclear spin system. These new features are predicted
not only for the case of a double quantum dots occupied by a pair of electrons but also
for the case of a single quantum dot occupied by a single electron or a pair of electrons.

\end{abstract}
\maketitle

\section{Introduction}
Quantum state control in solid systems is a challenging task due to the strong coupling of solid state
systems to environments in contrast with the atomic systems in which the coupling to environments is much weaker.
However, the prospect of realizing scalable architectures for quantum information processing motivated the
investigation on solid state/semiconductor structures.
Electron spins in semiconductor quantum dots(QD) proved to be one of the most promising two-level systems for
the quantum state control\cite{Loss98} due to their long decoherence times. Main decoherence mechanisms are the
coupling to phonons via the spin-orbit interaction and the hyperfine(HF) interaction with the
host nuclei.
The spin-orbit interaction leads to an exponential decay of the longitudinal and transverse electron spin
components characterized by $T_1$ and $T_2$ times\cite{Khaetskii00,Golovach04}. Under the strong confinement
and a weak magnetic field, the phonon-mediated decoherence is greatly suppressed,
whose time constant was demonstrated to reach up to $100$ms\cite{Meunier06b}. Instead
the contact hyperfine(HF) interaction of the electron spin with the lattice
nuclei dominates the decoherence\cite{Merkulov02,Coish05,Khaetskii03}.
 Contrary to the spin-orbit-mediated decoherence, the HF interaction can lead to the pure
dephasing and it features a Gaussian decay.
The HF interaction acts on a time scale proportional to the square root of the number of nuclei $\hbar/T_2^*=A/\sqrt{N}$,
where $A$ is the material specific HF coupling constant, and $N$ is the number of host nuclei. For example,
for GaAs $A=90\mu$eV\cite{Paget77} and for a QD having $10^6$ nuclei, the HF induced decoherence time $T_2^*$ is $\sim 10$ns.
In order to suppress the HF induced decoherence,
there have been made several proposals, such as the measurement of the HF field
\cite{Giedke05,Klauser05,Stepanenko06} and the polarization of nuclear spins, which
will reduce the fluctuations in the HF field\cite{Bracker05}.
 However, to achieve these, one has to do highly precise measurements or to polarize the nuclear spins to a high
degree.

 In small mesoscopic structures the
HF interaction is so far the only mechanism to probe nuclear spins, since typically NMR signals from such
small ensembles of nuclear spins is too weak to detect.
Coherent manipulation of mesoscopic ensemble of nuclear spins has been realized in semiconductor point
contact devices where magnetization of nuclear spins is probed by resistance measurements\cite{Yusa05}.
Hyperfine interactions lead to many interesting effects, such as lifting of spin blockade in transport through
double QDs\cite{Koppens05}, oscillatory currents in the spin blockade regime driven by the HF field\cite{Ono04},
and probing nuclear spin relaxation in the Coulomb blockade regime\cite{Huttel04}.
Coherent manipulation of the spin state of a pair of electrons on a double QD has been achieved via electrical
control of the exchange energy difference between the singlet and triplet spin states\cite{Petta05}, where
the singlet-triplet mixing via the HF interaction has been observed.

In light of these recent progress in the studies on the electron spin qubits and the HF interaction
in QDs, we are going
to investigate the quantum dynamics of the electron-nuclei coupled spin system, especially
the manipulation and preparation of nuclear spin states via the HF interaction,
which in turn lead to interesting effects, such as bunching in electron spin measurements and
the electron state revival\cite{Cakir07a,Cakir07b}. 

 Our paper is organized as follows.
In Sec.\,II, we are going to discuss a double QD model and the HF interaction and make comparison with available
experimental
data to derive relevant physical parameters. In Sec.\,III, we study the bunching in electron spin
measurements, which arises as a result of correlations between the successive electron spin measurements induced by the HF interaction. In Sec.IV, we will show that nuclear spins can be conditionally
purified via electron spin measurements, the manifestation of which is the revival of the electron spin state, enabling
the retrieval of an arbitrary electron spin state.
These newly predicted phenomena, bunching and revival, are not necessarily restricted to the case of an
electron pair in a double QD and can be observed in more general cases.
In Sec.\,V, we discuss the feasibility to observe these phenomena
in the electron spin measurements for a single QD occupied by either a single electron 
or a pair of electrons. 
Finally, our results and predictions are summarized in Sec.\,VI. 

\section{Double Quantum Dot Model}
We are going to consider a laterally coupled double QD system occupied by two electrons. QDs are formed on a  two-dimensional electron gas under a uniform magnetic field and the dynamics is assumed to take place only in the transverse spatial coordinates denoted by $x$ and $y$. In this section Zeeman energies are not taken into account because they are not essential for the orbital dynamics.  The orbital motion of electrons are governed by the Hamiltonian\cite{Burkard99,Taylor06}:
\begin{eqnarray}
H=\sum_{i=1,2} \{\frac{\bigl({\bf p}_i+\frac{e}{c} {\bf A}({\bf r}_i)\bigr)^2}{2m}+V(x_i,y_i) \}+V_c(|\mbox{\boldmath $\rho$}_1-\mbox{\boldmath  $\rho$}_2|) \label{full_hamilt}\\
V(x,y)=\frac{1}{2}\frac{m\omega^2}{4a^2}(x^2-a^2)^2+\frac{1}{2} m\omega^2 y^2 -\varepsilon x \;, 
V_C(r)=\frac{e^2}{\kappa r}\;,  \\
{\bf A}({\bf r})=\frac{B}{2}(-y,x,0)
\end{eqnarray}
where the confining potential is modeled by a double well potential which can be approximated by a harmonic potential near $x=\pm a$, the two-dimensional vector is represented by {\boldmath$\rho$}$=(x, y)$, $\varepsilon$ the external electric field , $e$ the elementary electric charge ($e>0$), $c$ the light velocity in vacuum and $\kappa$ is the dielectric constant.
Assuming the low temperature such that $\hbar\omega\gg kT$, we study the dynamics within the manifold of the ground state orbitals, consisting of $\dn{20,S}$, $\dn{11,T_{0,\pm}}$, $\dn{11,S}$ and $\dn{02,S}$. Here $\dn{nm, S(T_{\pm,0})}$ denotes the state with the electron occupation number $n(m)$ in the left(right) dot and $S$ and $T_{\pm,0}$ indicate respectively the singlet and triplet spin states. When both electrons are in the same dot, they are always in the singlet state since the orbital part is symmetric. However, when they are in different dots, the orbital part may be in an antisymmetric or a symmetric combination of ground state orbitals of the left and right QDs and thus the spin state may be a triplet or singlet state. For example, the orbital part of $\dn{11,S(T)}$ is given by
\begin{equation}
[\phi_L(1)\phi_R(2)\pm \phi_L(2)\phi_R(1)]/\sqrt{2}
\end{equation}
for the symmetric($+$)/antisymmetric($-$) combination of two electrons: one electron is localized in the left and the other in the right QD. The orbital state 
$\phi_L(1)\phi_R(2)$ is an eigenstate of the Hamiltonian (\ref{full_hamilt}) approximated by the local harmonic potentials excluding the Coulomb potential: 
\begin{eqnarray}
&& H_0=\sum_{i=1,2} \frac{\bigl({\bf p}_i+\frac{e}{c}{\bf A}({\bf r}_i)\bigr)^2}{2m}+V_0(x_1+a,y_1)+V_0(x_2-a,y_2)-\varepsilon (x_1+x_2) \;  \label{H0} \\
\mbox{with~}&& V_0(x,y)=\frac{1}{2}m\omega^2(x^2+y^2).
\end{eqnarray}
Then the ground eigenstate is given by  displaced harmonic oscillator states :
\begin{eqnarray}
\phi_L(1)\phi_R(2)=\phi(x_1+a-\frac{\varepsilon}{m\omega^2})\phi(y_1)\phi(x_2-a-\frac{\varepsilon}{m\omega^2})\phi(y_2) \:,  
\end{eqnarray}
where each wavefunction satisfies
\begin{eqnarray}
\bigl[\frac{\bigl({\bf p}+\frac{e}{c}{A}({\bf r})\bigr)^2}{2m}+\frac{1}{2}m\omega^2((x\pm a)^2+y^2))-\varepsilon x\bigr]\phi(x\pm a-\frac{\varepsilon}{m\omega^2})\phi(y)\nonumber \\=(\hbar\Omega\pm\varepsilon a-\frac{\varepsilon^2}{2m \omega^2})\phi(x\pm a-\frac{\varepsilon}{m\omega^2})\phi(y)\;,\\
\mbox{with}\quad \phi(x-x_0)=\frac{1}{\sqrt{\sqrt{\pi}\ell}}\exp[-\frac{(x-x_0)^2}{2\ell^2}]\exp[-ie x_0By/(2\hbar c)]\;,\\
 \ell^2=\frac{\hbar}{m\Omega}\;,\Omega=\sqrt{\omega^2+\frac{\omega_c^2}{4}}\;,\omega_c=\frac{eB}{mc}
\end{eqnarray}

For the orbital parts of $\dn{02,S}$ and $\dn{20,S}$ one has to calculate the eigenstates of two electrons occupying a single QD including the Coulomb potential:
\begin{eqnarray}
&& H_0=\sum_{i=1,2}\bigl\{\frac{\bigl({\bf p}_i+\frac{e}{c}{\bf A}({\bf r}_i)\bigr)^2}{2m}+\frac{1}{2}m\omega^2[(x_i\pm a)^2 +y_i^2]-\varepsilon x_i\bigr\}+\frac{e^2}{\kappa|\mbox{ \boldmath $\rho$}_1-\mbox{\boldmath $\rho$}_2|} \label{onsiteh} \;. 
\end{eqnarray}

When the onsite Coulomb energy is smaller than the orbital energy splitting, the $|02,S\rangle$ or $|20,S\rangle$ state orbital can be approximated by the product of the ground state orbitals of the harmonic oscillator. In this case, the ground state energy for the $|02,S\rangle$ and $|20,S\rangle$ state is given by
\begin{eqnarray}
E_{02}=2\hbar\Omega-2\varepsilon a-\frac{\varepsilon^2}{m\omega^2}+\delta_C\;, \; E_{20}=2\hbar\Omega+2\varepsilon a-\frac{\varepsilon^2}{m\omega^2}+\delta_C\;,
\end{eqnarray}
where the onsite Coulomb energy $\delta_C$ is calculated as 
\begin{eqnarray}
\delta_C=\sqrt{\pi/2}e^2/(\kappa \ell).
\end{eqnarray}


The exchange energy is found by calculating the energy of the $\dn{11,S(T)}$ state using the full Hamiltonian (\ref{full_hamilt}): 
\begin{eqnarray}
&&\langle{11,S(T)}|H\dn{11,S(T)}=\hbar\Omega + E_C\pm E_X\pm (t_{0R} \langle \phi_R|\phi_L\rangle+t_{0L} \langle \phi_L|\phi_R\rangle)\\
\mbox{with}\quad && E_C=\langle \phi_R(1)\phi_L(2)|\frac{e^2}{\kappa|{\bf r}_1-{\bf r}_2|}|\phi_R(1)\phi_L(2)\rangle\; ,\\
&& E_X=\langle \phi_R(1)\phi_L(2)|\frac{e^2}{\kappa|{\bf r}_1-{\bf r}_2|}|\phi_R(2)\phi_L(1)\rangle\;,\\
&& t_{0R}=\langle \phi_L |\delta V_R(x) |\phi_R\rangle\;, \delta V_R(x)=\frac{1}{2}\frac{m\omega^2}{4a^2}(x^2-a^2)^2-\frac{1}{2}m \omega^2(x-a)^2\;,\\
&& t_{0L}=\langle \phi_R |\delta V_L(x) |\phi_L\rangle\;, \delta V_L(x)=\frac{1}{2}\frac{m\omega^2}{4a^2}(x^2-a^2)^2-\frac{1}{2}m \omega^2(x+a)^2\;,
\end{eqnarray}
where $+(-)$ corresponds to the $\dn{S}(\dn{T})$ state, $E_C$ is the direct Coulomb energy,  $E_X$ the exchange integral and $t_{0R}$ is almost equal to $t_{0L}$ when $a\gg \varepsilon/(m\omega^2)$ and $t_0$ defined by $t_0=t_{0R}\simeq t_{0L}$ has a meaning of the single particle tunneling amplitude.
Restricting the Hamiltonian to the relevant two electron states, it is given as 
\begin{eqnarray}
&& H=\!\!2\hbar\Omega \mathds{1}+(-2\varepsilon a +\delta_C)\mat{02,S}{02,S}+ (2\varepsilon a +  \delta_C)\mat{20,S}{20,S}\nnb
&&\quad\quad +(E_C-\frac{j}{2})\sum_{\sigma=\pm,0}\mat{11,T_\sigma}{11,T_\sigma} +(E_C+\frac{j}{2})\mat{11,S}{11,S}\nnb
&&\quad\quad +t_R(\mat{11,S}{02,S}+ {\tt h.c.})+t_L(\mat{11,S}{20,S}+ {\tt h.c.})\quad\label{2elh} \\  \mbox{ with}&& \quad \quad
t_R=\ldn{11,S}\delta V_R(x_1)+\delta V_R(x_2)\dn{02,S}\;, \quad j=2E_X+4t_0\av{\phi_R|\phi_L},\nnb
&& \quad\quad t_L=\ldn{11,S}\delta V_L(x_1)+\delta V_L(x_2)\dn{20,S},
\end{eqnarray}
where $t_R$ and $t_L$ are the tunneling amplitudes and it can be shown that $t_R\simeq t_L\simeq t=\sqrt{2}t_0$,
when the onsite Coulomb energy is smaller than the orbital energy splitting and the $|02,S\rangle$ or $|20,S\rangle$ state orbital reduces to the product of the ground state orbitals of the harmonic oscillator. The structure of this Hamiltonian can be seen clearly in the matrix form:
\begin{eqnarray}
\left(\begin{array}{cccc}
\dn{20,S} & \dn{02,S} & \dn{11,S} & \dn{11,T_\sigma}\\
2\varepsilon a +\delta_C & 0 & t & 0\\
0 & -2\varepsilon a +\delta_C& t & 0\\
t & t & E_C+j/2 &0\\
0&0 &0 & E_C-j/2
\end{array}\right).
\end{eqnarray}

When the energy offset between the two QDs by the electrical bias is quite large, namely, $\varepsilon a\gg |t|$, one can consider the dynamics only in the $(1,1)$ and $(0,2)$ charge states, 
where the energy of the state $\dn{11,S}$ is renormalized by 
\begin{equation}
\delta E=t^2/(-2\varepsilon a+E_C+j/2-\delta_C)
\end{equation}
in consequence of the adiabatic elimination of the $(2,0)$ charge state. 

The Hamiltonian (\ref{2elh}) can be put in a simpler form:
\begin{eqnarray}
&& H=
-\Delta/2 \mat{02,S}{02,S}+\Delta/2 \mat{11,S}{11,S}+ t(\mat{11,S}{02,S}+ h.c.)  \nnb &&\quad \quad +(\Delta/2-j-\delta E)\sum_{\sigma=0,\pm}\mat{11,T_\sigma}{11,T_\sigma}, \;\\
\mbox{with} &&
\Delta=2\varepsilon a+E_C-\delta_C+j/2 +\delta E \;,\label{hsimp}
\end{eqnarray}
which is offset by some constant energy with respect to (\ref{2elh}).
$\dn{02,S}$ and $\dn{11,S}$ charge states hybridize to form new eigenstates $\dn{-,S}$ and $\dn{+,S}$ given by
\begin{equation}
\dn{\pm, S}=\frac{1}{\sqrt{(\Delta/2\mp \sqrt{\Delta^2/4+t^2})^2+t^2}}\bigl[t\dn{11,S} -(\Delta/2\mp \sqrt{\Delta^2/4+t^2})\dn{02,S}\bigr] \label{hybrid}
\end{equation}
and the Hamiltonian is rewritten as 
\begin{eqnarray}
H=
\sqrt{\frac{\Delta^2}{4}+t^2}\bigl[\mat{+,S}{+,S}-\mat{-,S}{-,S}\bigr] +(\frac{\Delta}{2}-j-\delta E)\sum_{\sigma=\pm,0}\mat{11,T_\sigma}{11,T_\sigma}.\label{hel}
\end{eqnarray}
When $\Delta\gg |t|$, $\dn{+(-),S}\rightarrow \dn{11(02),S}$, whereas when $\Delta$ is negative and $|\Delta|\gg |t|$, $\dn{+(-),S}\rightarrow \dn{02(11),S}$.
The energy difference between the singlet ground state and the triplet states is 
\begin{eqnarray}
J=\Delta/2-j-\delta E+\sqrt{\Delta^2/4+t^2}\label{exc}
\end{eqnarray}
and this energy will be called the "exchange energy" in the following. In this expression "$j+\delta E$"
term coming from the bare exchange integral and the level shift due to the transfer integral appears and its magnitude will be estimated from the comparison of $J$ with experimental data.
For vanishing external magnetic field the exchange energy should always be positive\cite{Mattis}, namely the ground state is always a singlet state. However, in the presence of a magnetic field, a singlet-triplet crossing takes place at some particular value of the magnetic field, yielding a triplet ground state\cite{Burkard99}, i.e., $J<0$.


\subsection{Hyperfine Interaction}
Now we are going to discuss the effects of the HF interaction with  nuclei. The HF interaction is mainly described by the Fermi contact interaction\cite{Abragam96}:
\begin{eqnarray}
V_{HF}=A \;v_0\sum_{i,\alpha}{\bf S}_i\cdot {\bf I}_\alpha\delta({\bf r}_i-{\bf R}_\alpha).
\end{eqnarray}
Here ${\bf r_i}$ denotes the position of the $i$th electron, and ${\bf R}_\alpha$ is the position of the nucleus $\alpha$. $A$ is a material specific coupling constant and for instance for GaAs $A=90\mu$eV and $v_0$ is the unit cell volume. $S$ and $I$ are the spin angular momenta of the electron and the nucleus, respectively. When two electrons are in the same QD, they experience the same HF field, which implies a vanishing HF field for singlet states(which is not the case for triplet states). On the other hand, when the electrons are in different QDs, the mean HF field induces mixing within triplet states, and the difference of the HF fields in two QDs  induces coupling between the singlet and triplet states. For two electrons in $\dn{11,T_{0,\pm}}$ and $\dn{11,S}$ states, the HF interaction is given as
\begin{align}
V_{HF}=&\frac{1}{2}(h_L+h_R)\cdot ({\bf S}_1+{\bf S}_2)+\frac{1}{2}(h_L-h_R)\cdot ({\bf S}_1-{\bf S}_2)\label{hf4d}\\
=&\frac{h_z}{2}(\mat{11,T_+}{11,T_+}-\mat{11,T_-}{11,T_-})\nnb&+\frac{1}{2\sqrt{2}}(h_-\mat{11,T_+}{11,T_0}+h_-\mat{11,T_0}{11,T_-}+{\tt h.c.})\nnb 
&\!\!+\frac{1}{2\sqrt{2}}(-\delta h_-\mat{11,T_+}{11,S}+\delta h_+\mat{11,T_-}{11,S}+{\tt h.c.})\nnb &+\frac{1}{2}(\delta h_z\mat{11,S}{11,T_0}+{\tt h.c.}) \\  &\mbox{with}\; {\bf h}={\bf h}_L+{\bf h}_R \;, \quad \delta{\bf h}={\bf h}_L-{\bf h}_R  \;,\quad
 {\bf h}_{L(R)}=A\; v_0\sum_\alpha |\phi_{L(R)}(R_\alpha)|^2{\bf I}_\alpha \;,
\end{align}
where ${\bf h}_{L(R)}$
is the HF field in the left(right) QD and has the dimension of energy. Thus the HF fields ${\bf h}$ and $\delta{\bf h}$ also have the dimension of energy.  In general the nuclear Zeeman energy is very small, for example, for $^{69}$Ga with $g_N=2.02$ it is about $0.74$mK at $B=1$T. Thus, for higher temperatures nuclear spins are randomly oriented and the HF field features 
a Gaussian distribution with the mean square value: 
\begin{equation}
\av{h^2_{L(R)}}=A^2\;v_0^2\sum_\alpha |\phi_{L(R)}(R_\alpha)|^4I(I+1),\label{rmsl} 
\end{equation}
where $I$ is the magnitude of the nuclear spin and $\av{\ldots}$ denotes the ensemble average. In particular, for the uniform coupling, we have 
\begin{equation}
\sqrt{\av{h_{L(R)}^2}}=A\; \sqrt{I(I+1)}/\sqrt{N_{L(R)}} \;,
\end{equation}
where $N_{L(R)}$ is the number of nuclear spins in the left(right) dot.

When the electron Zeeman energy is much larger than the HF fields, the coupling terms among the triplet states and those between $T_{\pm}$ and the singlet state $S$ can be neglected and the HF interaction reduces to
\begin{eqnarray}
V_{HF}=\delta h_z(\mat{11,S}{11,T_0}+{\tt h.c.})/2, \label{hfz}
\end{eqnarray}
with $\delta h_z=h_{Lz}-h_{Rz}$ being the difference of the HF fields along the applied field direction. All other spin states are unaffected by the HF interaction. 

Two electron system on a double QD is initialized in $\dn{02,S}$ state under the condition that  $\Delta\gg |t|$. If the bias voltage is changed adiabatically so that the singlet state remains in the ground state $\dn{-, S}$ without ever populating $\dn{+, S}$(\ref{hybrid}), the double QD electronic Hamiltonian(\ref{hel}) including the HF interaction (\ref{hfz}) is cast into the form:
\begin{align}
H+V_{HF}&=\frac{r\delta h_z}{2}(\mat{-,S}{11,T_0}+{\tt h.c.}) +\frac{J}{2}(\mat{11,T_0}{11,T_0}-\mat{-,S}{-,S}) \\
&=JS_z+r \delta h_z S_x \label{eq_hf}\\
&\mbox{with}\;\; \;r=t/\sqrt{(\Delta/2+\sqrt{\Delta^2/4+t^2})^2+t^2} \;,
\end{align}
where the factor $r=\av{11,S|-,S}$ determines the HF coupling strength of the singlet ground state(\ref{hybrid}) to the triplet state.
Now we examine the limiting values of $r$. The parameter $\Delta$ can be controlled by the bias voltage through $\varepsilon$ in (\ref{hsimp}) and $t$ can be varied through the spatial overlap of wavefunctions. When $\Delta$ is positive and $\Delta\gg |t|$, two electrons are almost localized in the right dot forming a spin singlet pair and $r\rightarrow 0$. On the other hand, when $\Delta$ is negative and $|t|\ll |\Delta|$, two electrons are separated in different dots with negligible spatial overlap. Then $|r|\rightarrow 1$ and the HF interaction is maximized. In Eq. (\ref{eq_hf}) the Hamiltonian is written in the pseudospin representation with 
$\dn{11,T_0}$ and $\dn{-,S}$ forming the bases.
   

\subsection{Singlet-triplet mixing}
Due to the HF interaction electrons prepared in the singlet state can be flipped to the triplet states. The spin state of an electron pair  evolves under the Hamiltonian (\ref{eq_hf}). The initial state of nuclear spins is assumed to be in an ensemble, where nuclear spins are randomly oriented. Then the time evolution of the density matrix of the electron pair-nuclei coupled system is given as
\begin{eqnarray}
\rho(t=0)=\sum_n p_n\hat{\rho}_n \mat{S}{S}\rightarrow
\rho(t)=\sum_n p_n\hat{\rho}_n \mat{\Psi_n(t)}{\Psi_n(t)}\nnb
\dn{\Psi_n(t)}= (\cos \frac{\Omega_n t}{\hbar}+i\frac{J}{2\Omega_n}\sin \frac{\Omega_n t}{\hbar}) \dn{S}  - i\frac{r h_n}{2\Omega_n} \sin \frac{\Omega_n t}{\hbar} \dn{T_0} \;, \quad \Omega_n=\sqrt{r^2h_n^2+J^2}/2 \;, \label{stevol}
\end{eqnarray}
where $\hat{\rho}_n$ characterizes the nuclear spin state which assumes the HF field value: $ \hat{\delta h}_z\hat{\rho}_n =h_n\hat{\rho}_n$, the weight of which is $p_n$, namely $\sum_n p_n=1$, and ${\rm Tr} \hat{\rho}_n=1$. In the following  $\hbar$ will be set to unity($\hbar=1$) for simplicity.
 From (\ref{stevol}) the probability to detect the triplet(singlet) state follows as
\begin{eqnarray}
&&P_T=\frac{1}{2}\av{\frac{r^2h^2}{ r^2h^2+J^2}(1-\cos\sqrt{r^2h^2+J^2}t)},\label{tripprob}\\
&& P_S=1-P_T(t),\label{prb2d} 
\end{eqnarray}
where $\av{\ldots}$ denotes the ensemble average over the HF fields.
When the nuclear spins are unpolarized and randomly oriented, the spectral weight of the HF field $p_n$ in (\ref{stevol}) follows a Gaussian profile\cite{Merkulov02}:
\begin{equation}
p[h]=\frac{1}{\sqrt{2\pi\sigma^2}}e^{-h^2/2\sigma^2} \;. \label{gaussian}
\end{equation}
This is the continuum expression under the correspondence of $p_n\rightarrow p[h]$. $\sigma^2 =\av{\delta h_z^2}$ is the mean square value of the HF field operator $\delta h_z$. Since the nuclear spins in the left and right dots are statistically independent, we have
\begin{equation}
\av{\delta h_z^2}=\av{( h_{Lz}-h_{Rz})^2}=\av{h_{Lz}^2}+\av{h_{Rz}^2} \;.
\end{equation}
Thus $\sigma^2$ is the sum of the mean square values of the HF fields in the left and the right QDs. 

For vanishing exchange coupling $J=0$ and $r=1$, (\ref{tripprob}) features a Gaussian decay:
\begin{eqnarray}
P_T=1/2(1-\exp[-\sigma^2t^2/2]) \;,
\end{eqnarray}
whereas for finite $J$, in the limit of $t\gg J/\sigma^2$, it shows a power law decay\cite{Coish05}:
\begin{eqnarray}
P_T=\frac{1}{2}\av{\frac{h^2}{h^2+J^2}}-\frac{\sqrt{2}}{\sqrt{J}\sigma t^{3/2}}\cos[Jt+3\pi/4]\label{psas}
\end{eqnarray}
which has been experimentally demonstrated\cite{Laird06}.
In the case  of a vanishing external magnetic field, all singlet and triplet states are coupled via the HF interaction(\ref{hf4d}).
 We consider the same situation, namely, an electron pair is initialized in the singlet state and after the HF interaction of duration $t$ the spin state of the electron pair is measured. Probability for singlet detection is given as
\begin{eqnarray}
 P_S=\Bigl\langle\langle S|1/4-{\bf S}_1(t)\cdot {\bf S}_2(t)|S\rangle\Bigr\rangle.\label{4dsingp}
\end{eqnarray}
The solution of (\ref{hf4d}) in the Heisenberg picture yields
\begin{eqnarray}
 {\bf S}_1(t)=\hat{h}_L\hat{h}_L\cdot {\bf S}_1+({\bf S}_1-\hat{h}_L\hat{h}_L\cdot {\bf S}_1)\cos h_Lt
+\hat{h}_L\times{\bf S}_1\sin h_Lt \label{4d1}
\end{eqnarray}
 with $\hat{h}_L={\bf h}_L/|{\bf h}_L|$ and its  ensemble average over ${\bf h}_L$ is calculated as
  \begin{eqnarray}
  \av{{\bf S}_1(t)}= \Bigl(1+2(1-\sigma_L^2t^2)\exp[-\sigma_L^2t^2/2]\Bigr){\bf S}_1/3.\label{4d2}
  \end{eqnarray}
Here $\sigma_L^2=\av{ {\bf h}_L^2}/3$
and similarly the expression for  ${\bf S_2}(t)$, i.e., $\av{{\bf S_2}(t)}$ is obtained by the replacements ${\bf h}_L\rightarrow {\bf h}_R$ and $\sigma_L\rightarrow \sigma_R$.
Using (\ref{4d1}) and (\ref{4d2}) the singlet detection probability(\ref{4dsingp}) can be readily evaluated: \cite{Schulten78}
\begin{eqnarray}
P_S=1/4+[1+2(1-\sigma_L^2t^2)e^{-\sigma_L^2t^2/2}][1+2(1-\sigma_R^2t^2)e^{-\sigma_R^2t^2/2}]/12\label{b0}
\end{eqnarray}
which yields  $ 1/3$ as $t\rightarrow\infty $.

\begin{figure}[h]
\includegraphics[width=\linewidth]{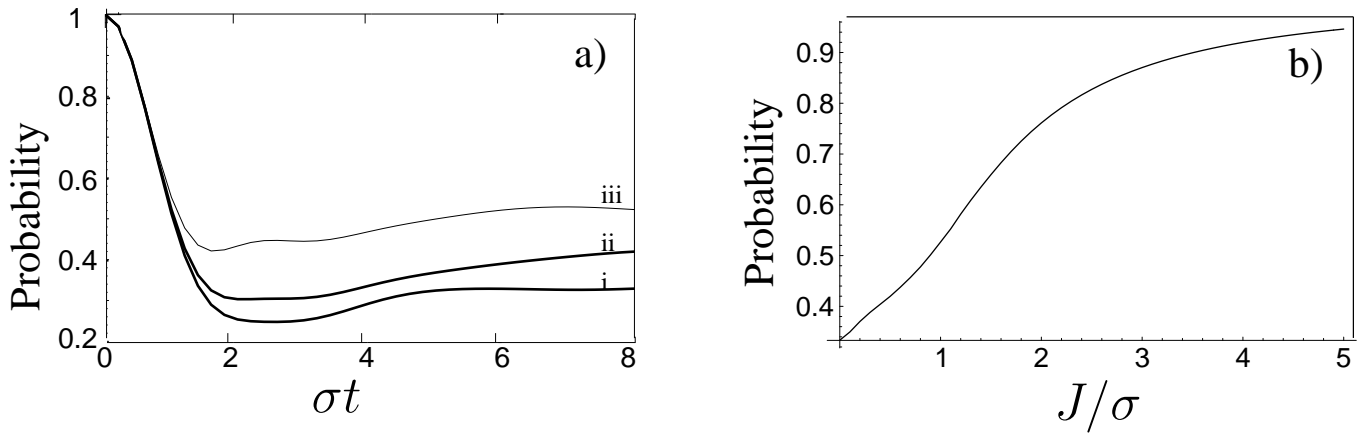}
\caption{ a)$P_S(t)$ (Eq.(\ref{prb4d})) as a function of time for $J/\sigma$ values of i) $0 $, ii) $0.5$, iii) $1$.
b)$P_S(t\rightarrow \infty)$ as a function of $J/\sigma$. These are calculated when no magnetic field is  applied.\label{figst4d}}

\end{figure}

When $J\neq 0$ and no magnetic field is applied, the Hamiltonian is as follows:
\begin{eqnarray}
H={\bf h}_L\cdot {\bf S}_1+{\bf h}_R\cdot {\bf S}_2+J({\bf S}_1\cdot{\bf S}_2+\frac{1}{4}) \;. \label{h4d}
\end{eqnarray}
Within the semiclassical model,
 we have diagonalized (\ref{h4d}) to find the singlet detection probability:
\begin{eqnarray}
P_S(t)=\bigl\langle\bigr|\sum_{i=1\ldots 4} |\ldn{S} e_j\rangle|^2 e^{-ie_j t}\,\bigl|^2\bigr \rangle,\label{prb4d}
\end{eqnarray}
where $|e_i\rangle,\,i=1,\ldots,4$ are the eigenvectors of (\ref{h4d}) for given ${\bf h}_{L}$ and ${\bf h}_{R}$ values.
 $\av{\ldots}$ denotes ensemble averaging over the HF fields featuring a Gaussian distribution (\ref{gaussian})
for ${\bf h}_{L(R)}$
assuming $\sigma_L=\sigma_R$ , i.e., $\sigma=\sqrt{\sigma_L^2+\sigma_R^2}=\sqrt{2}\sigma_{L(R)}$.
 The time dependence of $P_S(t)$ in (\ref{prb4d}) and the asymptotic value
 $P_s(t\rightarrow \infty)$ vs $J/\sigma$ are shown in Fig. \ref{figst4d}a) and Fig. \ref{figst4d}b), respectively.
 The singlet probability does not feature oscillations for finite $J$ values, in contrast to the case of
$|S\rangle-|T_0\rangle$ mixing (Eq. (\ref{psas})) which features oscillations at long time scales. This is due to
the destructive interference between contributions from four eigenstates.


\subsection{Comparison with experimental data}

\begin{figure}[h]
\includegraphics[width=\linewidth]{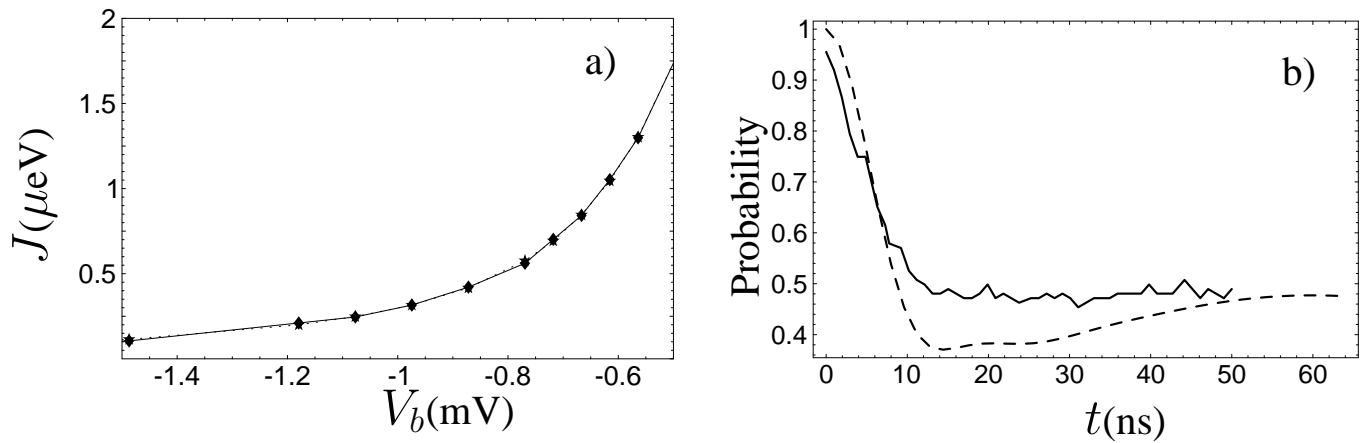}
\caption{a) Experimental values(dots) and theoretical fitting(solid line) for the exchange energy $J$ as a function
of the bias voltage $V_b$ which is a linear function of the detuning $\Delta$. b) The singlet detection probability
$P_S$ as a function of the duration time of the HF interaction without an external magnetic field; experiments in
Ref.[\onlinecite{Petta05}] (solid line) and theoretical results(dashed line). \label{fig_exp}}
\end{figure}

Now we make comparison of the above theoretical results with available experimental data to derive relevant physical parameters.
In the experiments by Petta et al.\cite{Petta05}, the exchange energy $J$ has been obtained as a function of the
bias voltage $V_b$ and their experimental data are shown in Fig.\ref{fig_exp}a) by dots.
We fitted the experimental data to the expression (\ref{exc}) assuming a linear relation between the bias voltage
$V_b$ and the detuning $\Delta/2=q V_b+E_0$. The expression for the exchange energy becomes
\begin{eqnarray}
J=-J_0+qV_b+E_0+\sqrt{(qV_b+E_0)^2+t^2}\label{fitj}
\end{eqnarray}
 which is fitted to experimental values yielding, $J_0=\delta E+j=0.07 \;\mu$eV, $q/e=5.86 \times 10^{-3}$,
$E_0=3.24 \;\mu$eV and $t=1.43\; \mu$eV. The fitting is performed in the range
$\Delta\in [-9.4,1]\mu{\tt eV}$ or equivalently $V_b\in [-2.15,-0.46]{\tt mV} $, which is exhibited in Fig. \ref{fig_exp}a)
with the solid line.  Due to an applied magnetic field, $B\sim 100$mT, the exchange energy (\ref{fitj}) can become negative for particular values of the bias voltage. Here the singlet-triplet crossing ($J=0$) occurs at a bias voltage $V_b=-5.53$mV. 

In the experiments \cite{Petta05} the singlet-triplet mixing data were obtained as a function of the HF interaction
period at the bias voltage $V_b=-6$ mV. For instance, the solid line in Fig. \ref{fig_exp}b) exhibits the
experimental probability to detect the singlet state
as a function of the HF interaction period, where no external magnetic field is applied(see Eq. \;(\ref{prb4d})). In the fitting procedure first we
determined the $J/\sigma$ ratio from the asymptotic value $P_S(t\gg 1/\sigma)$.
In particular, the experimental data shown by a solid line in Fig. \ref{fig_exp}b) exhibit an asymptotic value
$P_S(t\gg 1/\sigma)\sim 0.47$ which corresponds to the value $J/\sigma= 0.8 $ in Fig. \ref{figst4d}b).
At the next step, by matching the width of the main peak of the $P_S(t)$ profile for
$J/\sigma=0.8$ and the experimental data, the value of $\sigma\simeq 0.09\mu$eV is determined and thus
 $|J|\simeq 0.07\mu$eV is fixed.
On the other hand, we obtain
$J\simeq -0.038 \mu$eV when we insert $V_b=-6$mV in the expression (\ref{fitj}). 
This discrepancy by a factor about 2 may be induced by the inaccuracy of (\ref{fitj}) because $V_b=-6$mV is out of the range of fitting, i.e., [-2.15,-0.46]mV. However, the agreement in order of magnitude is rather satisfactory.  




\section{Bunching of Electron Spin Measurements}

Now that we have formulated the basic features of the electron-nuclei coupled system, we can examine the details of its quantum dynamics. First of all, we reveal an interesting phenomenon of  bunching of electron spin measurements which is caused by the correlation among successive measurements and is induced by the long-lived quantum coherence of nuclear spins. We also discuss the effect of relaxation of nuclear spins on this phenomenon of bunching.

\subsection{Successive measurements of electron spins} 
Now we show that by electron spin measurements in a double QD governed by the Hamiltonian in (\ref{eq_hf}), the coherent behavior of nuclear spins can be demonstrated. Corresponding to the experiments\cite{Petta05}, we assume that
 an electron pair is initialized in the singlet state and the nuclear spin states are initially in a mixture of  $\delta h_z$ eigenstates (\ref{stevol}).
In the unbiased regime, i.e., $r=1$, the nuclear spins and the electron spins interact for a time span of $\tau$.
  Then
the gate voltage is swept adiabatically, switching off the HF interaction, namely $r\rightarrow 0$, in a time scale much shorter than the duration time of the HF interaction $\tau$.
Next a charge state measurement is performed which detects the singlet or triplet state. Probability to detect the singlet or triplet state is calculated as
\begin{eqnarray}
&&P_S=\sum_n p_n|\alpha_n|^2, \quad P_T=\sum_n p_n|\beta_n|^2 \\
\mbox{with}&& 
\alpha_n=\cos\Omega_n \tau+iJ/2\Omega_n \sin\Omega_n \tau, \quad
\beta_n=-ih_n/2\Omega_n\sin\Omega_n \tau,
\end{eqnarray}
where the notations in (\ref{stevol}) are used.
Subsequently one can again initialize the system in the singlet state of the electron pair and  turn on the hyperfine interaction for a time span of $\tau$ and perform the second measurement.
In general over $N$ times measurements, the nuclear state conditioned on $k(\leq N)$ times singlet and $N-k$ times triplet detection is 
\begin{eqnarray}
\sigma_{N,k}=\bigl(^N_{\,k}\bigr)\sum_n p_n|\alpha_n|^{2k}|\beta_n|^{2(N-k)}\hat{\rho}_n \;, \label{nucstate}
\end{eqnarray}
the trace of which yields the probability to have $k$ times singlet outcomes:
\begin{align}
 P_{N,k}={\tt Tr}\sigma_{N,k}
 =\bigl(^N_{\,k}\bigr)\langle|\alpha|^{2k}|\beta|^{2(N-k)}\rangle,\label{Pqm}
\end{align}
where $\langle\ldots\rangle$ is the ensemble average over the HF field $h_n$ \cite{Merkulov02}.
 Hereafter, this case will be referred to as the {\it coherent regime}.
One can easily contrast this result with that for the {\it incoherent} regime in which nuclear spins  lose their coherence between the successive spin measurements and relax to the equilibrium distribution. The latter is given by
\begin{eqnarray}
P'_{N,k}=\bigl(^N_{\,k}\bigr)\langle|\alpha|^2\rangle^{k} \langle|\beta|^2\rangle^{(N-k)}.
\label{Psc}
\end{eqnarray}
When the nuclear spins are incoherent, the probability distribution (\ref{Psc}) obeys simply a Gaussian distribution with a mean value of $k=N\langle|\alpha|^2\rangle$ and the variance of   $N\langle|\alpha|^2 \rangle \langle|\beta|^2\rangle$, as $N\rightarrow\infty$.
 However, when nuclear spins preserve their coherence, the probability distribution (\ref{Pqm}) may exhibit different statistics depending on the initial nuclear state.
 The two probability distributions (\ref{Pqm}) and (\ref{Psc}) yield the same mean value, i.e., $
\overline{k}=N\langle|\alpha|^2\rangle$,
but with distinct higher order moments. If the weight factor $p_n$ of the HF field in the equilibrium distribution  has a width $\sigma$, then for the duration time of the HF interaction  $\tau\geq 1/\sigma$, the distributions (\ref{Pqm}) and (\ref{Psc}) start to deviate from each other.
 They yield the same distribution only when the initial nuclear state is in a well defined eigenstate of $\delta h_z$, i.e., when $\sigma=0$.

If the nuclear spins are coherent over the span of the experiment, then
successive electron spin measurements are biased to all singlet(triplet) outcomes.
In particular, when the initial nuclear spins are unpolarized and randomly oriented, the distribution of the hyperfine field is characterized by a Gaussian distribution (\ref{gaussian}) with the variance $\sigma^2$.
As the simplest case, let us check the results of two measurements, each following the HF interaction of duration $t$. Probabilities in the coherent and incoherent regimes for two singlet detections are respectively calculated as
\begin{eqnarray}
P_{2,2}=\langle|\alpha|^4\rangle=\{6+2 e^{-2\sigma^2 t^2}+8e^{-\sigma^2 t^2/2}\}/16\nnb
P'_{2,2}=\langle|\alpha|^2\rangle^2=\{4+8e^{-\sigma^2 t^2/2}+4e^{-\sigma^2 t^2}\}/16 \;,
\end{eqnarray}
where results are given particularly for $J=0$ and it turns out that  $P_{22}>P'_{22}$ . 


\begin{figure}[!t]
\includegraphics[width=\linewidth]{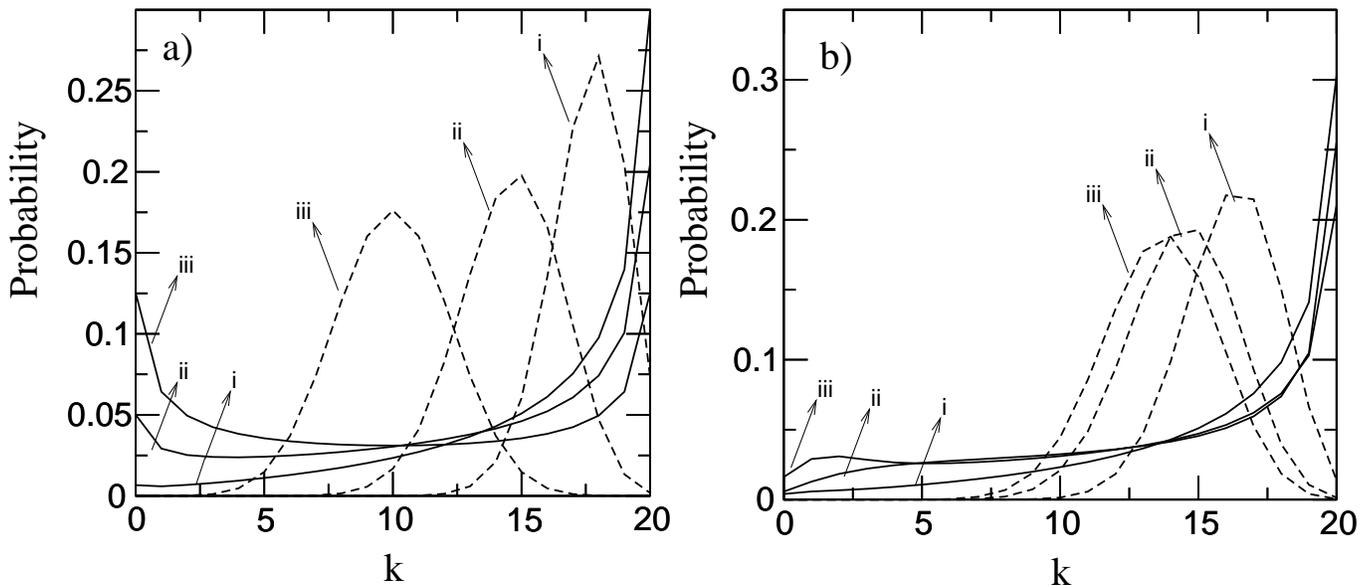}
\caption{Probability distribution $P_{Nk}$  at $N=20$ measurements for $k={0,1,\ldots,20}$ times singlet detections, for coherent regime(solid lines) and incoherent regime(dashed lines). Two cases of the exchange energy are considered  a) $J=0$ b) $J/\sigma=0.5$ for HF interaction periods of $\sigma\tau=$ i)$0.5$, ii)$1.5$, and iii)$\infty$. \label{Fig_20meas}}
\end{figure}

In Fig. \ref{Fig_20meas}, for $N=20$ measurements, $P_{N,k}$ is shown
 for three values of the duration time of the HF interaction: $\sigma\tau=0.5, 1.5, \infty$. For $\tau=0$, the probability for both (\ref{Pqm}) and (\ref{Psc}) is peaked at
$k=20$. However, immediately after the HF interaction is introduced, the probability distributions show distinct behaviors.  The measurement results in the incoherent regime approach a Gaussian distribution. On the other hand, in the coherent case, the probabilities
bunch at $k$=0 and 20 for $J=0$ and when $J/\sigma=0.5$ those bunch at $k=20$ only. As $J$ is increased above some critical value, no bunching takes place at $k=0$ times singlet measurement, since the singlet state becomes energetically stable and the state change to the triplet state becomes unfavorable.

To observe the bunching $N$ successive spin measurements are performed within the coherence time of the nuclear spins. Then after waiting for some time so that nuclear spins are again randomized, another set of
$N$ successive measurements are carried out and so on. Thus an ensemble average of $N$ measurements is performed which results in a bunching of either spin singlet or triplet outcomes. This bunching is a clear signature of coherent behavior of nuclear spins, which can easily be contrasted with the incoherent regime which merely exhibits a Gaussian distribution.

\subsection{Effects of nuclear spin diffusion}
Now we  will discuss the effect of nuclear spin diffusion on the bunching of electron spin measurements, which leads to a transition from the coherent regime to the incoherent regime.
During the interval  between the successive measurements the nuclear spin state relaxes to the equilibrium distribution due to the dipole-dipole interactions \cite{Abragam96}. 

If the substrate surrounding a QD is of the same kind of material as that of the QD, nuclear states will diffuse due to the interaction with the surrounding nuclei, which leads to a change both in the total spin angular momentum of the nuclei and the HF field. The inhomogeneous distribution of HF coupling constants will also induce a redistribution of the spin angular momentum leading to a change in the HF field.
Since a detailed discussion on the nuclear spin diffusion is beyond the scope of this paper, we simply
 develop a phenomenological argument based on the diffusion equation in the phase space of the HF field:
\begin{eqnarray}
\frac{\partial p[h,t]}{\partial t}=\kappa \frac{\partial^2}{\partial h^2}(p[h,t]-p_0[h]) \:, \label{diff}
\end{eqnarray}
where $p_0[h]$ is the distribution corresponding to the steady state configuration of nuclear spins.   At high temperatures compared with the nuclear Zeeman splitting, $p_0[h]$ obeys a Gaussian distribution(\ref{gaussian}). 
 The general solution of the diffusion equation (\ref{diff}) can be cast into the form:
\begin{eqnarray}
p[h,t]=\frac{1}{\sqrt{2\pi\sigma^2}}e^{-\frac{h^2}{2\sigma^2}}-\frac{1}{\sqrt{2\pi(\sigma^2+2\kappa t)}}e^{-\frac{h^2}{2(2\kappa t+\sigma^2)}}
 +\frac{1}{\sqrt{4\pi \kappa t}}\int{\tt d}h'e^{-\frac{(h-h')^2}{4\kappa t}}p[h',t=0] \;,  
\end{eqnarray}
where $p[h,t=0]$ is the initial distribution of the HF field. 
 
 The randomization of nuclear spins will lead to loss of memory effects described in the last section. The 
 nuclear state conditioned on the electron spin measurements (Eq. (\ref{nucstate})) will decohere throughout  the successive measurements. The nuclear diffusion time is much longer than the characteristic time  of the spin singlet-triplet mixing induced by the HF interaction (Eq. (\ref{eq_hf})), namely  $t_{diff}\gg \tau\sim 1/\delta h_z$. Decoherence of nuclear spins will mainly take place during the electron spin measurement, because this process of spin-charge conversion is time-consuming \cite{Petta05}.
 
 For instance, when an electron pair is initialized in the spin singlet state starting with a randomized nuclear spin configuration (\ref{gaussian}), then subject to the hyperfine interaction (\ref{eq_hf}) for $J=0$, of duration $\tau$ and is followed by the electron spin measurement, the spectrum of the  HF field becomes  
 \begin{eqnarray}
 p[h]={\cal N}e^{-h^2/2 \sigma^2}(1\pm \cos h\tau) \;, \label{dist}
 \end{eqnarray}
corresponding to either singlet(+) or triplet(-) outcome, where ${\cal N}$ is a normalization constant.
Governed by the diffusion equation (\ref{diff}), the distribution (\ref{dist})  after a time span of $T$ evolves to 
 \begin{eqnarray}
 p[h,T]= \frac{1}{\sqrt{2\pi\sigma^2}}e^{-\frac{h^2}{2\sigma^2}} +
\frac{e^{-\frac{h^2}{2(2\kappa T+\sigma^2)}}}{\sqrt{2\pi(2\kappa T+\sigma^2)}}\Bigl \{\frac{ 1\pm e^{-\frac{\sigma^2\tau^2 \kappa T}{2\kappa T+\sigma^2}}\cos \frac{h \tau}{1+2\kappa T/\sigma^2}}{1\pm e^{\frac{-\sigma^2\tau^2}{2}}}-1\Bigr\} .\label{nucdiffmeas}
 \end{eqnarray}
This distribution converges to a Gaussian for $T\gg \sigma^2/\kappa$.  It is also to be noted that the duration
of the HF interaction $\tau$ also affects the effective diffusion time. This means that the period of
modulation $\simeq 1/\tau$ induced in the nuclear field spectrum affects the speed of diffusion.
 In fact (\ref{nucdiffmeas}) approaches the Gaussian form $p_0[h]$ when $\tau^2\gg 1/\kappa T,1/\sigma^2$,
namely when the period of undulation in the nuclear field spectrum is short enough to be smoothed out easily,
the distribution converges to $p_0[h]$.

\begin{figure}[]
\includegraphics[width=\linewidth]{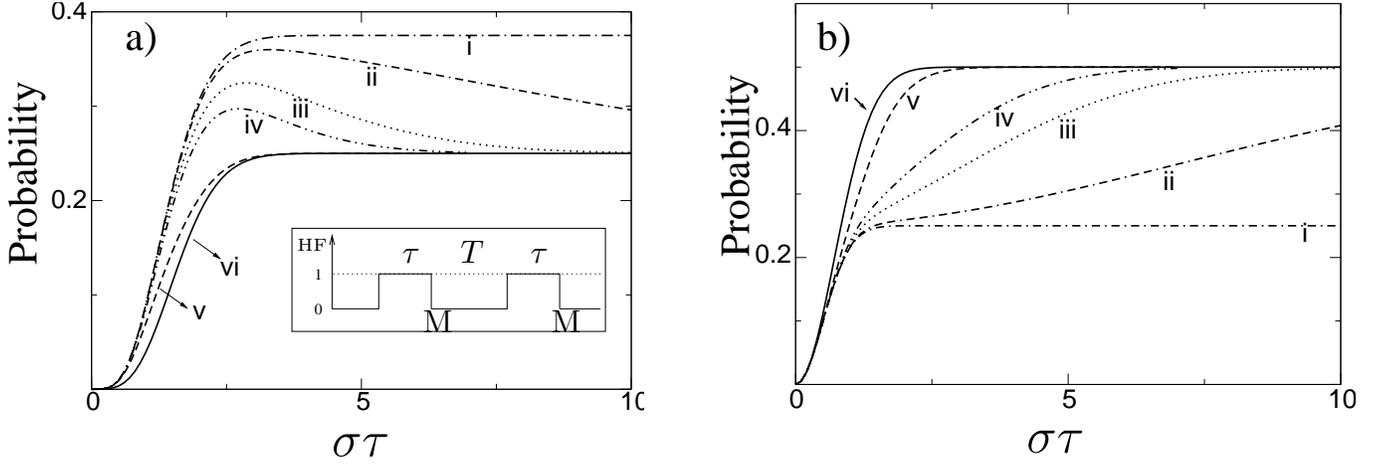}
\caption{Results of two successive spin measurements: Probability of  a) two triplet b)one singlet-one triplet measurements, at $\kappa T/\sigma^2=$i)$0$, ii)$0.01$, iii)$0.05$, iv)$0.1$, v)$0.5$,  and vi) $\infty$. Inset: (Measurement scheme)  $\tau$ is the duration of the HF interactions, and $T$ the waiting time with HF interaction switched off, where electron spin measurements are denoted by $M$.
 \label{Fig_nucdiff}}
\end{figure}

 After a time span of $T$ following the first measurement, the system is again initialized, HF interaction is
switched on for a time span $\tau'$, then a second spin measurement is performed. Here typically
$T\gg \tau,\tau'$. The measurement results approach that of semiclassical picture when
$T\gg \sigma^2/\kappa$ or $\tau^2\gg 1/\kappa T,1/\sigma^2$. Otherwise one can still trace the nuclear
memory effects in the measurement results. In Fig. \ref{Fig_nucdiff}, some examples are
shown for two successive measurements with parameter values of $\kappa T/\sigma^2=0,0.05,0.1,0.5,\infty$,
as a function of the HF interaction time $\tau=\tau'$. In the asymptotic limit of $\kappa T/\sigma^2=\infty$,
we can check that the probability in Fig. \ref{Fig_nucdiff} a)(b)) approaches $P'_{2,0}(P'_{1,1})$. 

\section{Purification of nuclear spin state and electron spin revivals}

\begin{figure}
\includegraphics[width=\linewidth]{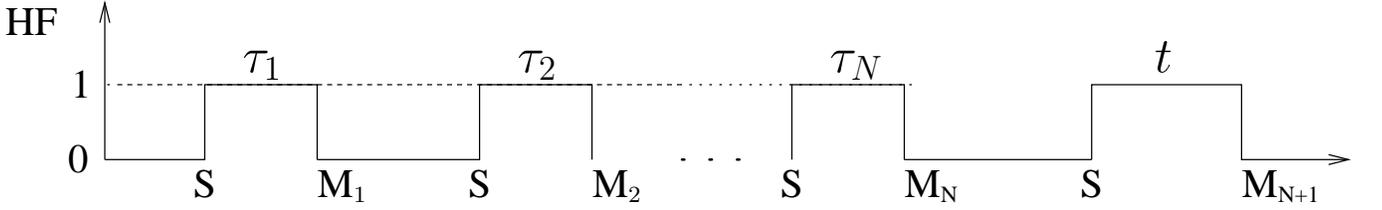}
\caption{Measurement scheme to observe the electron spin revival: Each time the electron spin is
initialized in the spin singlet state(denoted by S), then the HF interaction is switched on
for a period $\tau_i$ followed by the electron spin measurement with an outcome M$_i$, for $i=1,
\cdots, N.$ These are the preparation stage. Then after the HF interaction for a period $t$,
the $N+1$-th measurement is carried out. \label{fig_pur}  }
\end{figure}

In this section we investigate the conditional preparation and purification of nuclear spin state via
successive electron spin measurements.
This feature becomes manifest via revival phenomena of the electron spin state.
Here the HF interaction is assumed to take place in the unbiased
regime of the double QD, i.e., when $J=0$ and $r=1$ in (\ref{stevol}).
Then the nuclear state prepared by $N$ successive electron spin measurements
with $k$ times singlet outcomes, each following the HF interaction of duration times
$\tau_{1},\tau_2,\ldots ,\tau_N$ is given by
\begin{eqnarray}
\sigma_{N,k}={\cal N}\sum p_n\hat{\rho}_n \prod_{i=1}^k \cos^2\frac{h_n \tau_i}{2}\prod_{j=k+1}^N
\sin^2\frac{h_n \tau_j}{2} \;, \label{nucstate2}
\end{eqnarray}
where ${\cal N}$ is a normalization constant. The sequence of measurements is depicted in Fig. \ref{fig_pur}.
In the following we consider the case where all measurement outcomes are singlets and examine
two typical cases: A. $\tau_1=\tau_2=\ldots=\tau_N$ and B. $\tau_1=2\tau_2=\ldots=2^{N-1}\tau_N$.

\subsection{ First case: $\tau_1=\tau_2=\ldots=\tau_N=\tau$}

\begin{figure}
\includegraphics[width=\linewidth]{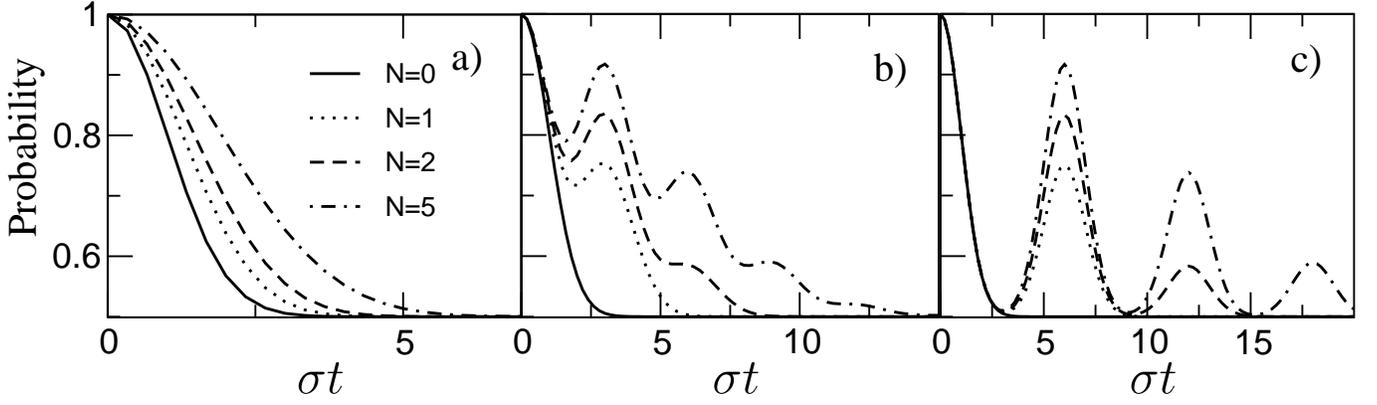}
\caption{Conditional probability for singlet state detection as a function of HF interaction period
$\sigma t$, subject to $N=0,1,2,5,10$ times prior singlet state measurements and for HF interaction duration times
a) $\sigma\tau=1.0$, b) $\sigma\tau=3.0$, and c) $\sigma\tau=6.0$. \label{fig-cond}}
\end{figure}

In this case all the duration times of the HF interaction are equal and the prepared nuclear state
following $N$ times singlet measurements is given by
\begin{eqnarray}
\hat{\sigma}={\cal N}\sum_n p_n\hat{\rho}_n \cos^{2N}\frac{h_n\tau}{2} \;, \label{condstate}
\end{eqnarray}
where ${\cal N}$ is a normalization constant.
 Given the initial state $\hat{\rho}(t=0)=\hat{\sigma}\mat{S}{S}$, the probability to measure the
singlet electron spin state after the HF interaction of duration time $t$ is calculated as
\begin{align}
P(t;\lbrace\tau_i=\tau\rbrace_{i=1,\ldots,N})&=\frac{\av{\cos^{2N}[h\tau/2]\cos^2[ht/2]}}{\av{\cos^{2N}[h\tau/2]}}\nnb
 &=\frac{1}{2}+\frac{1}{4}\frac {\sum_{s=0}^{2N} \sum_{\alpha=\pm } \binom{2N}{s}
\av{\exp[i(s-N)h\tau+i\alpha h t]}} {\sum_{s=0}^{2N}\binom{2N}{s}\av{\exp[i(s-N)h\tau]}} \;, \label{deriv1}
\end{align}
where $\av{\ldots}$ denotes ensemble averaging with respect to the initially random nuclear spin
state(\ref{gaussian}). Using the identity:
\begin{eqnarray}
\av{e^{iht}}=\frac{1}{\sqrt{2\pi\sigma^2}}\int{\tt d}h e^{-h^2/2\sigma^2} e^{iht}=e^{-\sigma^2t^2/2} \;, \label{ident}
\end{eqnarray}
the equation (\ref{deriv1}) can be cast into the form:
\begin{eqnarray}
P(t;\lbrace\tau_i=\tau\rbrace_{i=1,\ldots,N})=\frac{1}{2}+\frac{1}{2}\frac{\sum_{s=0}^{2N}\binom{2 N}{s}e^{-\sigma^2(t-(N-s)\tau)^2/2}} {\sum_{s=0}^{2N}\binom{2N}{s}e^{-(s-N)^2\sigma^2\tau^2/2}}. \label{condprob1}
\end{eqnarray}
For $t <1/\sigma$ this gives a Gaussian decay(see Fig. \ref{fig-cond}a), whereas for
$t >1/\sigma$ it exhibits revivals(see Fig. \ref{fig-cond}c).
For $\sigma \tau \gg 1$ the expression(\ref{condprob1}) reduces to
\begin{eqnarray}
P(t;\lbrace\tau_i=\tau\rbrace_{i=1,\ldots,N})=\frac{1}{2}+\frac{1}{2}\frac{\sum_{s=0}^{2N}\binom{2 N}{s}e^{-\sigma^2(t-(N-s)\tau)^2/2}}
{\binom{2N}{N}} \;, \label{condproblarget}
\end{eqnarray}
featuring revivals at $t=n \tau (n=1,2,\ldots)$ with a decreasing amplitude:
\begin{equation}
1/2+\binom{2N}{N-n}/2\binom{2N}{N}
\end{equation}
which becomes $1/2+e^{-n^2/N}/2$ for $N\gg 1$.
In the method proposed here, the nuclear spin state can be conditionally purified without determining
the precise value of the HF field. Although the HF field may be still assuming indefinite values,
electron-nuclei correlations lead to revivals at known times.
As an example consider the case when the nuclear spin state is prepared by five HF interaction stages,
each of which has a duration time $\tau=10/\sigma$ and is followed by a singlet detection of the
electron spin state. This conditionally prepared nuclear spin state revives the spin singlet electron
 state at times $\sigma t=10,20,\ldots,50$ with fidelities $1/2+\binom{2N}{N-s}/2\binom{2N}{N}$,
for $N=5$ and $s=1,2,\ldots,5$ which are $11/12,31/42,\ldots,253/504$. Success probability to prepare
such a state is $\sim 1/2^5$.

In order to understand the physics of the revival more clearly, we consider the limit $N \gg 1$.
Since $|\cos \theta| \leq 1$, $\cos^{2N} \theta$ is sharply peaked at $\theta=s \pi$
($s \in \mathbb{Z}$) and can be approximated as
\begin{eqnarray}
&&\cos^{2N} \theta = \sum_{ s \in \mathbb{Z}} (1 - \frac{1}{2} (\theta- s \pi)^2+\cdots)^{2N} \nnb
&&=\sum_{ s \in \mathbb{Z}} (1 - N (\theta- s \pi)^2+\cdots) \simeq
\sum_{ s \in \mathbb{Z}} \exp[- N (\theta- s \pi)^2] \;.
\end{eqnarray}
Then the spectrum of the nuclear HF field corresponding to the conditionally prepared state (\ref{condstate})
can be approximated in the limit $N\gg 1$ as
\begin{eqnarray}
&&p[h] ={\cal N} e^{-h^2/2\sigma^2} \cos^{2N}\frac{h\tau}{2}
 \simeq {\cal N} e^{-h^2/2\sigma^2} \sum_{s\in \mathbb{Z}} e^{-(h-h_s)^2/2\sigma_m^2} \label{dist1} \\
{\rm with} && \sigma_m^{-1}=\tau \sqrt{N/2} \;,
\end{eqnarray}
implying squeezing of the HF spectrum at particular known values $h_s=2s\pi/\tau$.
Given the initial nuclear spin state with the spectrum (\ref{dist1}), the probability to recover an
initial singlet electron spin state after the HF interaction of duration time $t$ is given by
\begin{eqnarray}
P(t)=\av{ \cos^2{ht/2}}=
\frac{1}{2}+\frac{\cal N}{2}
 \; \sum_{s\in \mathbb{Z}}  e^{-\sigma_m^2t^2/2}e^{-h_s^2/2\sigma^2} \cos h_st \;, \label{probsing}
\end{eqnarray}
where it is assumed that $\sigma\gg \sigma_m=1/(\tau\sqrt{N/2})$ and the normalization constant
${\cal N}$ is set to satisfy $P(t=0)=1$.
 In (\ref{probsing}) each $h_s$ gives rise to revivals at times $t_s=\tau/2s$ and its integer multiples.
At the common multiples of all $t_s$ values which are $n\tau, n=1, 2, 3\ldots$, the probabilities add
up coherently leading to revivals(c.f. (\ref{condprob1})) and each revival has the amplitude
$1/2+e^{-n^2/N}/2$.

The revival phenomenon also applies to some arbitrary initial electron spin state subject to the HF
interaction with the conditionally prepared nuclear spin state (\ref{condstate}). When the initial
state of the system is assumed as
\begin{eqnarray}
\rho(t=0)=\hat{\sigma}\mat{\psi}{\psi},\quad \dn{\psi}=\cos\frac{\theta}{2}\dn{S}+\sin\frac{\theta}{2}
e^{-i\phi}\dn{T} \;, \label{initarb}
\end{eqnarray}
the fidelity $F=\av{\psi|\rho(t)|\psi}$ to recover the initial electron spin state $\dn{\psi}$ at
time $t$ is calculated as
\begin{equation}
F=\sin^2\theta\cos^2\phi+(1-\sin^2\theta\cos^2\phi)P(t) \label{fidelity}
\end{equation}
with $P(t)$ given by (\ref{deriv1}).

\subsection{Second case: $\tau_1=2\tau_2=\ldots=2^{N-1}\tau_N=\tau$}

 Here the duration times of the HF interaction are decreased by one half successively, namely $\tau=\tau_1=2 \tau_2=2^2\tau_3=\ldots=2^{N-1}\tau_N$. In this case the prepared state (\ref{nucstate2}) is given as
\begin{eqnarray}
\hat{\sigma}={\cal N} \sum_n  p_n\hat{\rho}_n \prod_{s=1}^N\cos^{2}\frac{h_n \tau}{2^s} \label{condstate2nd}
\end{eqnarray}
and the  probability to recover the initial singlet electron spin state after the HF interaction of duration time $t$ is calculated as
\begin{eqnarray}
P(t;\lbrace\tau_i=2^{-i+1}\tau\rbrace_{i=1,\ldots,N})=\frac{1}{2}+\frac{1}{2}\frac{\sum_{s_i=0}^2 \binom{2}{s_1} \binom{2}{s_2} \cdots \binom{2}{s_N} e^{- \sigma^2 (t-\sum_{i=1}^N (s_i-1)\tau/2^{i-1})^2/2}}{\sum_{s_i=0}^2 \binom{2}{s_1} \binom{2}{s_2} \cdots \binom{2}{s_N} e^{- \sigma^2 (\sum_{i=1}^N (s_i-1)\tau/2^{i-1})^2/2}} \;. \label{condprob2}
\end{eqnarray}
The singlet state is revived at $t=2\tau \times 0.l_1l_2\ldots l_N$, where $l_i=0,1$. This amounts to
$2^N$ revivals at times  $t=\tau/2^{N-1}, 2\tau/2^{N-1}, 3\tau/2^{N-1},\ldots, 2\tau(1-1/2^N), 2\tau$.

Now we  briefly discuss the HF spectrum of the state (\ref{condstate2nd}) and its relation to revivals in
(\ref{condprob2}). The HF spectrum of (\ref{condstate2nd}) can be cast into the form:
\begin{eqnarray}
p[h]={\cal N}e^{-h^2/2\sigma2} \frac{1}{4^N}\frac{\sin^2h\tau}{\sin^2\frac{h\tau}{2^{N}}} \;. \label{meas_2}
\end{eqnarray}
In the same way as in (\ref{dist1}), in the limit of $N \gg 1$, we can show that
\begin{eqnarray}
&&p[h]= {\cal N} e^{-h^2/2\sigma^2} \sum_{s\in \mathbb{Z}} e^{-(h-h_s)^2/2\sigma_m^2} \label{scheme2}\\
{\rm with}&&
h_s=s 2^N \pi/\tau \;\; {\rm and} \quad \sigma_m=\sqrt{\frac{3}{2}} \frac{1}{\tau} \;.
\end{eqnarray}
This implies squeezing of the spectrum at $h_s=s2^{N}\pi/\tau$, $s\in\mathbb{Z}$.
 Given the initial nuclear spin state with the spectrum (\ref{scheme2}), the probability to recover
the initial singlet electron spin state at time $t$ is given by the same expression as in
(\ref{probsing}).  Each $h_s$ leads to revivals at times $\tau/(s 2^N)$ and their integer multiples,
which add up coherently at $n\tau/2^{N-1}, (n=1, 2,\cdots)$, giving rise to revivals
(c.f. (\ref{condprob2})). Thus we can understand that the revival phenomena occur reflecting the
undulation in the nuclear field spectrum induced by the electron spin measurements.

As a concrete example we  make a comparison of the two schemes for $N=2, k=2$ and examine the electron spin revivals for this conditionally prepared state.
For $\tau_1=2\tau_2=\tau\gg 1/\sigma$,  the conditional
probability (\ref{condprob2}) is given as
\begin{eqnarray}
P(t;\lbrace\tau_1=\tau,\tau_2=\frac{\tau}{2}\rbrace)\simeq \frac{1}{2}+\frac{1}{8}\bigl\{ e^{-\frac{\sigma^2 (t-3\tau/2)^2}{2} }+
2e^{-\frac{\sigma^2(t-\tau)^2}{2} }+ 3e^{-\frac{\sigma^2(t-\tau/2)^2}{2} }+
4e^{-\frac{\sigma^2 t^2}{2} } \bigr\} \;, \label{cond2}
\end{eqnarray}
whereas for $\tau_1=\tau_2=\tau\gg 1/\sigma$, (\ref{condprob1}) is calculated as
\begin{eqnarray}
P(t;\lbrace\tau_1=\tau,\tau_2=\tau\rbrace)\simeq\frac{1}{2}+\frac{1}{12}\bigl\{ e^{-\frac{\sigma^2(t-2\tau)^2}{2} }+
4e^{-\frac{\sigma^2(t-\tau)^2}{2} }+6e^{-\frac{\sigma^2 t^2}{2} }
\bigr\}.\label{cond3}
\end{eqnarray}
We have more revivals with higher probabilities for the former case in which the undulation in the
nuclear field spectrum is more structured.

 In the above we found that the nuclear field spectrum is squeezed or undulated through the electron
spin measurements. In order to examine the degree of squeezing quantitatively, we estimate the
purity of the nuclear spin system.
The purity of the system is given by ${\cal P}_{N,k}={\tt Tr} \; \sigma_{N,k}^2$. Using the identity ${\tt Tr} \; \hat{\rho}_n^2=1/{\cal D}p_n$, where $\cal D$ is the total dimension of the Hilbert space of nuclear spins, we obtain the purity of the state (\ref{nucstate2}) as
\begin{align}
{\cal P}_{N,k}&=\frac{1}{\cal D}\frac{\int {\tt d}h\,p[h] \prod_{i=1}^{k}\cos^4\frac{h\tau_i}{2}\prod_{j=k+1}^N\sin^4\frac{h\tau_j}{2}}{ \bigl( \int {\tt d}h\,p[h]\prod_{i=1}^{k}\sin^2\frac{h\tau_i}{2}\prod_{j=k+1}^N\sin^2\frac{h\tau_j}{2}\bigr)^2 } \label{purint}\\
&=\frac{1}{\cal D}\frac{\sum_{s_i=0}^
4(^{~4}_{s_1})(^{~4}_{s_2})\ldots(^{~4}_{s_N})e^{-\frac{1}{2}\bigl[(s_1-2)\taut_1+(s_2-2)\taut_2+\ldots
+(s_N-2)\taut_N\bigr]^2}(-1)^{s_{k+1}+\ldots+s_N}}{\bigl[ \sum_{s_i=0}^
2(^{~2}_{s_1})(^{~2}_{s_2})\ldots(^{~2}_{s_N})e^{-\frac{1}{2}\bigl[(s_1-1)\taut_1+(s_2-1)\taut_2+\ldots
+(s_N-1)\taut_N\bigr]^2}(-1)^{s_{k+1}+\ldots+s_N}\bigr]^2 } \;, \label{purity}
\end{align}
where $\taut=\sigma\tau$ and sums were evaluated in the continuum limit  $\sum_np_n\rightarrow\int {\tt d}h\,p[h] $ in (\ref{purint}). 
We can extremize the purity(\ref{purity}) by choosing appropriate duration times of the HF interaction . In the asymtotic limit $\taut_i\gg 1$, we have
\begin{eqnarray}
{\cal P}_{N,k}=\frac{1}{\cal D}\frac{\sum_{s_i=0}^
4\binom{4}{s_1}\ldots\binom{4}{s_N}(-1)^{s_{k+1}+\ldots+s_N}\delta[(s_1-2)\taut_1+\ldots+(s_N-2)\taut_N] }
{\bigl[ \sum_{s_i=0}^2 \binom{2}{s_1}\ldots\binom{2}{s_N}(-1)^{s_{k+1}+\ldots+s_N}\delta[(s_1-1)\taut_1
+\ldots+(s_N-1)\taut_N] \bigr]^2 }.\label{purityas}
\end{eqnarray}
From (\ref{purityas}) we see that there are several asymptotic values determined by the
roots of the linear equations $\sum(s_i-2)\tau_i=0$ and $\sum(s_i-1)\tau_i=0$. 

\begin{figure}
\includegraphics[width=\linewidth]{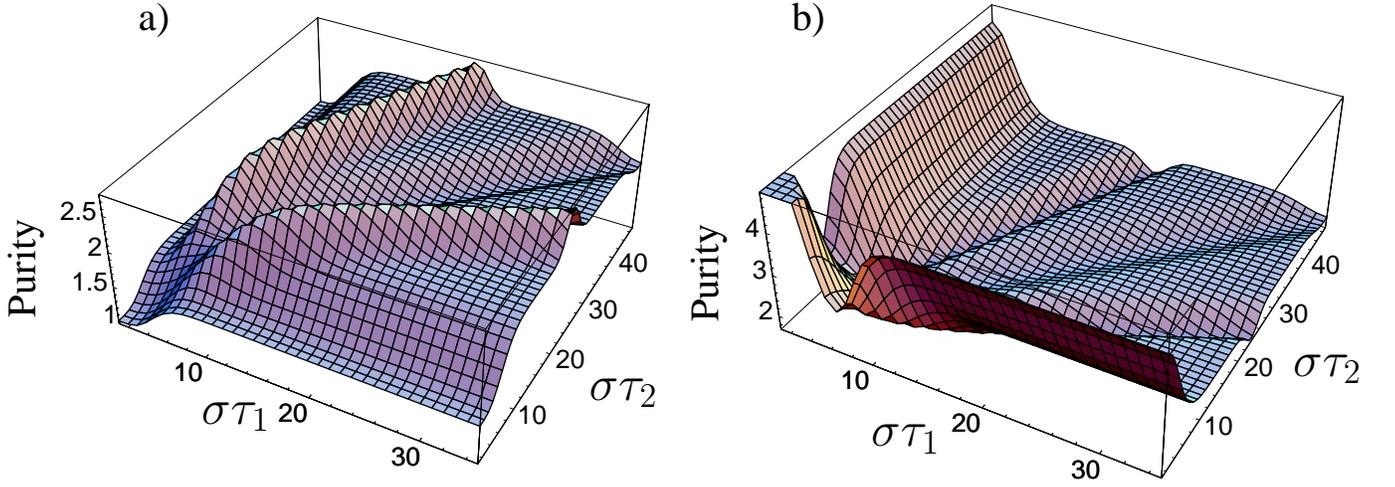}
\caption{(Color online)Purity(in units of $1/{\cal D}$) as a function of the duration times $\tau_1$ and $\tau_2$ of the HF interaction in the case of $N=2$ measurements for a)$k=2$ and b)$k=0$ times singlet outcomes.\label{fig_3dpur} }
\end{figure}

For instance in the case of $N=2$ and $k=2$, we see that the numerator in (\ref{purityas}) has a contribution from the choice of $s_1=s_2=2$, whereas the denominator has a contribution from the choice of
$s_1=s_2=1$, irrespective of the relative magnitude of $\tau_1$ and $\tau_2$. Additionally, in the case of $\tau_1=2\tau_2$ or $\tau_2=2\tau_1$ the numerator has finite contributions arising from some combinations of $s_1$ and $s_2$. On the other hand, the denominator does not have such contributions, because the equation $\sum_i (s_i -1)\tau_i=0$ cannot be satisfied except for $s_1=s_2=1$. In the case of $\tau_1=\tau_2$, both the numerator and the denominator have finite contributions from appropriate choices of $s_1$ and $s_2$ other than the trivial ones given by $s_1=s_2=2$ or $s_1=s_2=1$. Summarizing,
there are three asymptotic
limits (see Fig. \ref{fig_3dpur}a ), namely when i) $\tau_1=2\tau_2$ then ${\cal P}_{2,2}=11/4{\cal D}$,
ii) $\tau_1=\tau_2$ then ${\cal P}_{2,2}=35/18{\cal D}$ and c)otherwise ${\cal P}_{2,2}=9/4{\cal D}$.
In general the purity attains its maximum for all singlet outcomes, i.e., for $k=N$ and under the condition that the duration times of the HF interaction are halved at each step, viz., $\tau_i/\tau_{i+1}=2$.

\begin{figure}
\includegraphics[width=\linewidth]{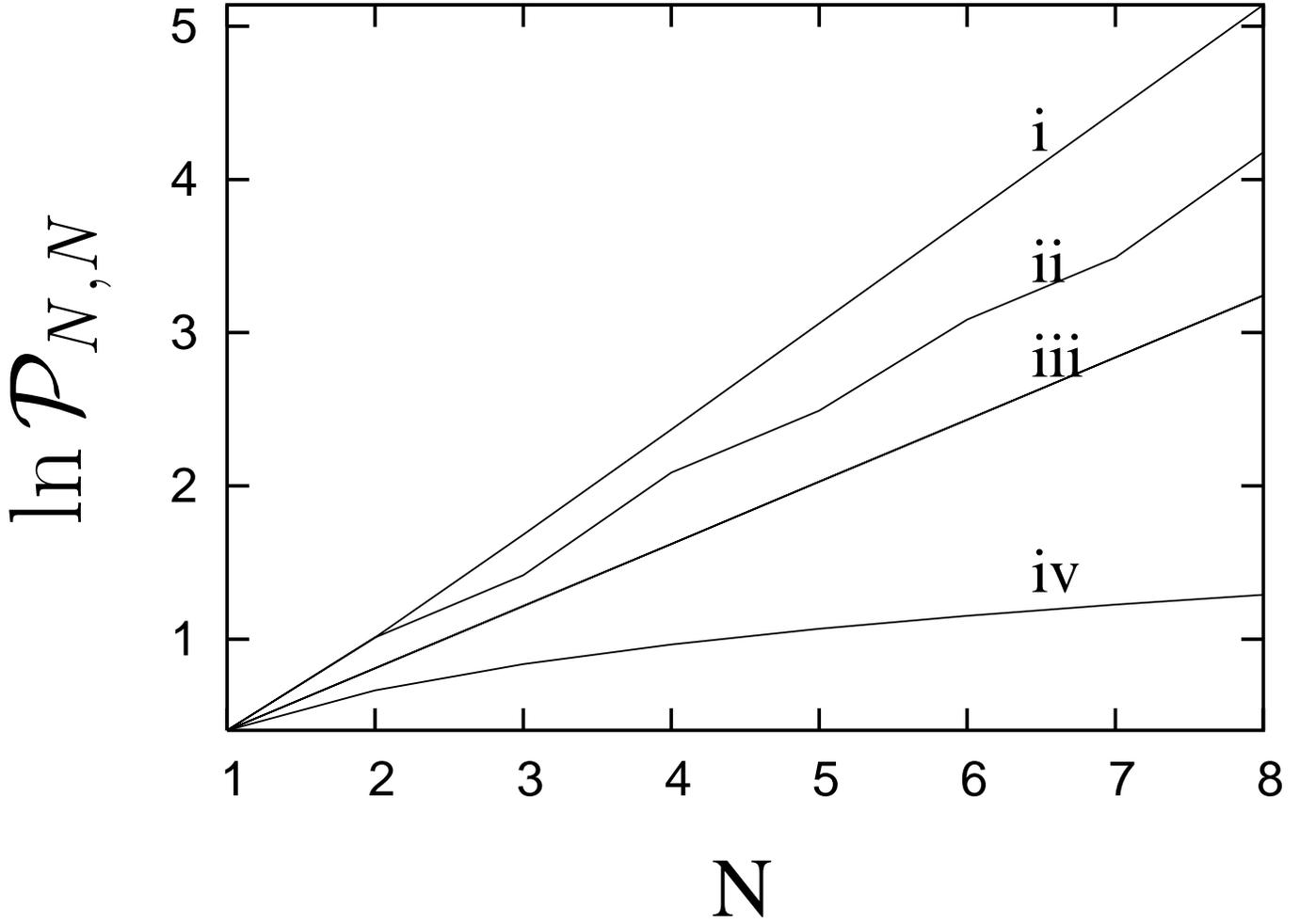}
\caption{Purity ${\cal P}_{N,N}$(in units of $1/\cal{D}$) is shown in the natural {\it logarithmic} scale,  for the schemes i)$\tau_1=2\tau_2=\ldots=2^{N-1}\tau_N$, ii)$\tau_1=2\tau_2=3\tau_3=\ldots=N\tau_N$, iii)$\tau_i:\tau_{j}$ is irrational for any pair of $(i,j)$,  and for iv) $\tau_1=\tau_2=\ldots=\tau_N$ as a function of the number of electron spin measurements($N=1,2,\ldots,8$).\label{purn}}
\end{figure}

In Fig.\,\ref{purn} the purity ${\cal P}_{N,N}$ is shown as a function of the number of measurements $N$, in the asymptotic limit of $\tau_i\gg 1/\sigma$, $i=1,\ldots,N$.  
The curve  i) corresponds to the maximum purity and the curve  iv) to the minimum, whereas all other choices of interaction periods $\tau_1:\tau_2:\ldots:\tau_N$ yield intermediate values(see Appendix \ref{purityform}). 


\section{Realization on a single QD}

\subsection{Single electron on a single QD}

So far we have discussed the bunching and revival phenomena only for a double QD system. The same predictions
can be made also for a single QD occupied by a single electron\cite{Hanson03,Dutt05,Atature06}.
Consider a single QD occupied by a single electron, under an external magnetic field such that the electron
Zeeman energy is much greater than the HF energies. Then the system is described by the Hamiltonian:
\begin{eqnarray}
H\simeq g_e\mu_BB S_z + h_z S_z =({   B_e}+h_z) S_z \;, \label{sqdh}
\end{eqnarray}
where $g_e$ is the electron $g$-factor, $\mu_B$ the Bohr magneton, $B$ the external magnetic field applied
in the $z$ direction, ${   B_e}$ represents the electron Zeeman energy and $h_z$ is the nuclear HF
field in
the $z$ direction . Spin flips are suppressed since ${   B_e}=g_e\mu_B B \gg
\sqrt{\langle {\bf h}^2\rangle}$.
The spin eigenstates in the $x$ direction $|\pm\rangle=(|\uparrow\rangle\pm|\downarrow\rangle)/\sqrt{2}$
are coupled by the HF interaction with $|\uparrow(\downarrow)\rangle$
being the eigenstates of $S_z$. The time evolution of the state $|+\rangle$ is simply given by ($\hbar=1$)
\begin{equation}
e^{-i Ht} \; |+\rangle = \cos \frac{{   B_e}+h_z}{2} t |+\rangle -i \sin \frac{{   B_e}+h_z}{2} t |-\rangle \;.
\end{equation}
Now let us consider the following experiment.
Each time the electron is prepared in the state $|+\rangle$. Next it is loaded into the QD, then removed from
the QD after some dwelling time $\tau$ and the spin measurement is
performed in the basis of $|\pm\rangle$. Essentially the same predictions
as those for a double QD can be made for this system, namely the electron spin bunching and revival.
We  consider the electron spin revival as an example.
{   After $N$ times the HF interaction of duration time $\tau$, each
followed by
 the measurement outcome of the $|+\rangle$ state,} the nuclear spin state becomes
\begin{eqnarray}
\hat{\sigma}=\sum_n p_n\hat{\rho}_n \cos^{2N}\frac{({   B_e}+h_n)\tau}{2} \;,
\end{eqnarray}
where the initial distribution $p_n$ characterizes the random distribution of the HF field (\ref{gaussian}).
Then the electron spin state is prepared in $\dn{+}$, yielding the initial state $\rho(t=0)=\hat{\sigma} \mat{+}{+}$ which evolves under the Hamiltonian (\ref{sqdh}).
The probability for obtaining $|+\rangle$ after the HF
interaction of duration $t$ is given as
\begin{align}
P(t;\lbrace\tau_i=\tau\rbrace_{i=1,\ldots,N})&=\frac{\av{\cos^{2N}\frac{({   B_e}+h)\tau}{2}\cos^{2}\frac{({   B_e}+h)t}{2}}}
{\av{\cos^{2N}\frac{({   B_e}+h)\tau}{2}}}\nnb
&=\frac{1}{2}+\frac{1}{4}\frac{\sum_{\alpha=\pm} {   \sum_{s=0}^{2N}} \binom{2N}{s}
\av{e^{i({   B_e}+h)[(s-N)\tau+\alpha t] }}}
{\sum_{s=0}^{2N}\av{\binom{2N}{s}e^{i(s-N)({   B_e}+h)\tau}}} \;, \label{ara}
\end{align}
where $\av{\ldots}$ denotes ensemble averaging with respect to (\ref{gaussian}).
Using the identity (\ref{ident}) the equation (\ref{ara}) for $\tau\gg 1/\sigma$ can be put in the form:
{  
\begin{align}
P(t;\lbrace\tau_i=\tau\rbrace_{i=1,\ldots,N})\simeq 1/2+\frac{1}{2(^{2N}_{~N})}\sum_{s=0}^{2N}(
^{2N}_{~s})e^{-\sigma^2(t-(N-s)\tau)^2/2}\cos B_e [t-(N-s)\tau]\;.
\end{align}
}
Thus, in a single QD, revivals are present as in the double QD case(c.f. Eqs. (\ref{condprob1}) and
(\ref{condproblarget})).


\subsection{A pair of electrons on a single QD}

The Hamiltonian (\ref{eq_hf}) can also be used to describe a pair of electrons in a single QD
\cite{Fujisawa02,Hanson05} and the same predictions as those for a double QD can be made. In the two electron
regime, the energy splitting between the singlet ground state and the triplet excited state can be tuned down
to zero by application of a magnetic field\cite{Hanson05,Meunier06b,Kyriakidis02} leading to  a singlet-triplet
crossing.
Under a high magnetic field, the triplet state ($T_0$) having zero magnetic quantum number is coupled to
the singlet state ($S$) via the HF field:
\begin{eqnarray}
h=A v_0\sum_i \phi_g({\bf R}_i) \phi^*_e({\bf R}_i)I^{(i)}_z/\sqrt{2} \;, \label{hf_sqd}
\end{eqnarray}
where $\phi_{g(e)}$ is the ground(excited) state orbital in the QD and the derivation is given
in Appendix A.
Typically for a two-dimensional QD with harmonic confinement, the HF field (\ref{hf_sqd}) has a mean square
value:
\begin{eqnarray}
\langle h^{\dag} h \rangle=A^2 I(I+1)v_0/16 \pi d r_0^2 \;,\label{hfsq}
\end{eqnarray}
where $r_0=\sqrt{\hbar/m \Omega}$ is the Fock-Darwin radius, $d$ the thickness of the QD,
$\Omega=\sqrt{\omega_0^2+\omega_c^2/4}$ with $\omega_0(\omega_c)$ being the frequency of the harmonic confinement
potential(the cyclotron frequency) and $I$ is the magnitude of the nuclear spin.
In the energy spectrum of a single QD the singlet-triplet crossing was observed via the tuning of
magnetic field\cite{Kyriakidis02}.

For an isotropic GaAs QD with a harmonic confinement energy $\omega=1$meV, a
 singlet-triplet crossing will take place at
$B\simeq 1.1$T and the second excited state which is a singlet, is separated by $\sim 0.2$meV.
   For such a QD, with thickness $d=1$nm, the rms value for the HF field (\ref{hfsq}) will be
$\sqrt{\langle h^{\dag} h \rangle}\sim 0.04\; \mu$eV, which implies that the system can be treated as a
two level system coupled by the HF field(See Appendix \ref{ap_spect} for the spectrum of a single QD occupied by two electrons).
The relevant Hamiltonian describing the dynamics within the subspace formed by $|S\rangle$ and
  $|T_0\rangle$ is
essentially the same as that for an electron pair in a double QD. Furthermore,
the electrons' spin state can be initialized and measured with high fidelity by a spin-selective coupling to
leads, relying on the spin-dependent tunnel rates\cite{Hanson05}.
Thus the observation of the same phenomena as the bunching in the electron spin measurements and the revival
of the initial electron state is feasible also in a single QD occupied by a pair of electrons.

\section{Conclusion}
We have investigated the quantum dynamics of the electron-nuclei coupled spin system in QDs and predicted some
interesting new phenomena. The quantum correlation induced in the system via consecutive HF interactions leads
to the bunching of outcomes in the electron spin measurements and the revival of an arbitrary initial
electron spin state. Simultaneously, the nuclear spin system is affected by the quantum correlation and is
in fact squeezed as confirmed by the increase in the purity. It is suggested that the consecutive electron spin
measurements provide a probabilistic method to squeeze or prepare the nuclear spin system. We also discussed
the effect of nuclear spin relaxation on the bunching and revival phenomena based on a phenomenological model
and exemplified a change from the coherent regime to the incoherent regime.
All the results obtained are applicable not only to a double QDs occupied by a pair of electrons but also to
a single QD occupied by a single electron or a pair of electrons, whenever the HF interaction is present and the nuclear spin state is
coherent throughout the experiments. 

\begin{acknowledgments}
We would like to thank Professor H. Kosaka for stimulating discussions and continual encouragements. This work is financially supported by the Japan Science and Technology Agency and also by the Ministry of Education, Culture, Sports, Science and Technology.
\end{acknowledgments}

\appendix
\section{Hyperfine interaction for an electron pair in a single QD}
Here we derive the Hamiltonian for an electron pair in a single QD. Under a sufficiently strong magnetic field,
the triplet states $T_{\pm}$ are well separated from the $T_0$ state and the singlet state $S$. Thus the
Hamiltonian within the subspace spanned by $T_0$ and $S$ states will be considered.
The wavefunctions for the $S$ and $T_0$ states are given, respectively, as
\begin{eqnarray}
&&\Psi_S({\bf r}_1, \xi_1, {\bf r}_2, \xi_2)=\phi_g({\bf r}_1) \phi_g({\bf r}_2)
\frac{1}{\sqrt{2}}(\alpha(\xi_1)\beta(\xi_2)-\beta(\xi_1)\alpha(\xi_2)) \;, \\
&&\Psi_{T_0}({\bf r}_1, \xi_1, {\bf r}_2, \xi_2)=\frac{1}{2}(\phi_g({\bf r}_1) \phi_e({\bf r}_2)
-\phi_e({\bf r}_1) \phi_g({\bf r}_2))(\alpha(\xi_1)\beta(\xi_2)+\beta(\xi_1)\alpha(\xi_2)) \;,
\end{eqnarray}
where $\phi_g(\phi_e)$ is the ground(excited) state orbital in the QD and $\alpha(\beta)$ denotes the spin
up(down) state. The HF interaction for two electrons is given by
\begin{equation}
V_{HF}=A v_0 \sum_i {\bf S}_1\cdot {\bf I}_i \; \delta({\bf r}_1-{\bf R}_i)
+ A v_0 \sum_i {\bf S}_2\cdot {\bf I}_i \; \delta({\bf r}_2-{\bf R}_i) \;,
\end{equation}
where ${\bf R}_i$(${\bf I}_i$) denotes the position(spin vector) of a nucleus and ${\bf S}_1$ and ${\bf S}_2$ are the electron spin vectors. 
Then we find
\begin{eqnarray}
&&\langle \Psi_{T_0}|V_{HF}|\Psi_{S}\rangle= -\frac{1}{\sqrt{2}} A v_0 \sum_i \phi_e^*({\bf r}_i)
\phi_g({\bf r}_i) I_{iz} \;, \; \langle \Psi_{S}|V_{HF}|\Psi_{T_0} \rangle=
\langle \Psi_{T_0}|V_{HF}|\Psi_{S} \rangle^* \;, \\
&& \langle \Psi_{S}|V_{HF}|\Psi_{S}\rangle = \langle \Psi_{T_0}|V_{HF}|\Psi_{T_0}\rangle =0\;.\label{matel}
\end{eqnarray}
Thus the singlet-triplet mixing is induced by the HF interaction. The effective nuclear
field operator coupling the singlet and triplet states in (\ref{matel}) will be introduced by
\begin{align}
h&=\langle \Psi_{T_0}|V_{HF}|\Psi_{S} \rangle \\ 
&= -\frac{1}{\sqrt{2}} A v_0 \sum_i \phi_e^*({\bf R}_i) \phi_g({\bf R}_i) I_{iz}
\end{align}
which has the dimension of energy and its mean square value is estimated as
\begin{eqnarray}
&&\langle h h^{\dag}\rangle=\frac{(A v_0)^2}{2}  \sum_i |\phi_e^*({\bf R}_i) \phi_g({\bf R}_i)|^2
\langle I^2_{iz} \rangle \\
&&=\frac{A^2 v_0}{2} \frac{I(I+1)}{3} \int d^3 r \; |\phi_e^*({\bf r}) \phi_g({\bf r})|^2 \;,
\end{eqnarray}
where $I$ is the magnitude of the nuclear spin.
Employing the envelope functions for the ground and excited states given by
\begin{eqnarray}
&&\phi_g(r, \theta, z)=\frac{1}{\sqrt{\pi}r_0} e^{-\frac{r^2}{2r_0^2}} \sqrt{\frac{2}{d}} \cos(\frac{\pi z}{d})\;, \\
&&\phi_e(r, \theta, z)=\frac{1}{\sqrt{\pi}r_0^2} e^{-\frac{r^2}{2r_0^2}} \; r e^{-i \theta} \sqrt{\frac{2}{d}}
\cos(\frac{\pi z}{d}) \\
{\rm with} && r_0=\sqrt{\frac{\hbar}{m\Omega}} \;, \; \Omega=\sqrt{\omega_0^2+(\frac{eB}{2mc})^2} \;,
\end{eqnarray}
where $d$ is the thickness of the QD, we have
\begin{equation}
\langle h h^{\dag}\rangle=\frac{A^2 v_0}{16 \pi r_0^2 d} \; I(I+1)\;.
\end{equation}

\section{ Energy spectrum of two electrons in a QD with isotropic harmonic confinement\label{ap_spect}}
\begin{figure}[!h]
\includegraphics[width=\linewidth]{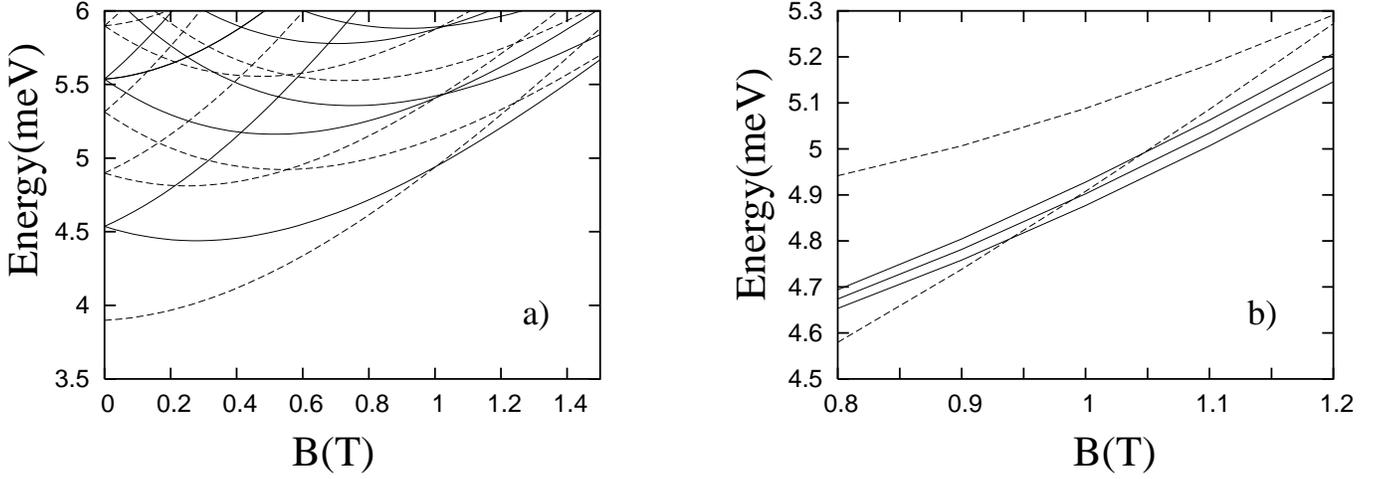}
\caption{a)Energy spectrum of two electrons in a GaAs-like QD  with a harmonic confinement energy
$\omega_0=1$meV.
Solid(dashed) lines indicate the triplet(singlet) states.
b)Energy spectrum in the vicinity of the singlet-triplet crossing. Triplet states $T_\pm$ are split by the Zeeman energy $\pm g\mu_B \simeq \mp 25.5 \mu{\tt eV/T}$ from the $T_0$ state as shown by solid lines. \label{specom1}}
\end{figure}

Here we calculate the energy spectrum of two electrons in a QD, assuming an isotropic harmonic confinement of
frequency $\omega_0$ in the $xy$ plane and a strong confinement along the growth($z$) direction.
Introducing the center-of-mass and the relative coordinates, the total
Hamiltonian can be divided as
\begin{align}
&H=H_{{\rm CM}}+H_{{\rm rel.}}+H_{\rm Z} \;, \\
&H_{{\rm CM}}=\frac{{\bf P}^2}{2M}+\frac{1}{2}M\Omega^2 {\bf R}^2 +\frac{\omega_c L^{\rm(CM)}_z}{2} \;, \\
&H_{{\rm rel.}}= \frac{{\bf p}^2}{2\mu}
+\frac{1}{2}\mu\Omega^2 {\bf r}^2 +\frac{\omega_c L^{\rm (rel)}_z}{2}+\frac{e^2}{\kappa r},\label{ham2d}\\
& H_{\rm Z}=g\mu_B B(S^{(1)}_z+S^{(2)}_z)\\
\mbox{with}\quad &{\bf R}=({\bf r}_1+{\bf r}_2)/2,\,\,{\bf P}={\bf p}_1+{\bf p}_2 \;,\,\,\,\,
{\bf r}={\bf r}_1-{\bf r}_2 \;,\,\, {\bf p}=({\bf p}_1-{\bf p}_2)/2, \\
&L_z^{\rm (CM)}={\bf R}\times {\bf P} \;,\,\,  L_z^{\rm (rel)}={\bf r}\times {\bf p} \;,\,\,\,
\Omega=\sqrt{\omega_0^2+\omega_c^2/4}\;,
\end{align}
where ${\bf r}_{1(2)}$ denotes the coordinates of the first(second) electron in the $xy$ plane,
$M=2m^*$, $\mu=m^*/2$ with $m^*$ being the electron effective mass, $\omega_c=eB/m^* c$  the cyclotron
frequency and $\mu_B$ is the Bohr magneton.
Employing $m^*=0.067 \cdot m_e$, $\kappa=12.53$ and $g=-0.44$\cite{Bornstein} appropriate for GaAs,
we diagonalized numerically the Hamiltonian $H_{{\rm rel.}}$ for the relative coordinate part. 
In the numerical diagonalization, 20 Fock-Darwin basis functions are employed to guarantee sufficient accuracy.
The energy spectrum for a GaAs-like QD with $\omega_0=1$meV is depicted in Fig. \ref{specom1}a) as a function of the
magnetic field $B$  for the orbital part $H_{\rm CM}+H_{\rm rel}$.
In Fig. \ref{specom1}b) the energy spectrum is plotted in the vicinity of the lowest energy singlet-triplet crossing point including the spin degrees of freedom. The singlet ground state and the triplet first excited states feature a crossing at $B\sim 1$T separated from
the next excited state by $\sim 0.2 $meV which is a singlet.
For a magnetic field $B<0.9$T, the electrons can be loaded into the singlet ground state and then by sweeping the magnetic field to the $S-T_0$ crossing point,  the system can be initialized. Here the $S-T_+$ crossing point should be passed at a rate much faster than the HF interaction time which was estimated as $\sim 103$ ns from $\sqrt{\av{h h^\dagger}}\sim 0.04\mu$eV in Sec.\,V-B. At the $S-T_0$ crossing point the $T_{+}$ and $T_-$ states are separated by a Zeeman energy $\sim 25\mu$eV which is much greater than the HF interaction energy. Thus at the $S-T_0$ crossing point the system can be described
 by a two level Hamiltonian composed of $\dn{S}$ and $\dn{T_0}$ states.
 As demonstrated in the experiments by Meunier et al.\cite{Meunier06b},  the phonon mediated spin relaxation time exceeds well beyond ms order which leaves the HF interaction as the only relevant mechanism at time scales shorter than ms.

\section{ Purity of the nuclear spin state as a function of the number of electron spin measurements
\label{purityform} }

When all $\tau_i:\tau_j, (i\neq j)$ ratios are irrational, in (\ref{purityas}), only $s_1=s_2=\ldots=s_N=2$ in the
numerator and only $s_1=s_2=\ldots=s_N=1$ in the denominator contribute, yielding
${\cal P}_{N,N}=(3/2)^N/{\cal D}$  (Fig.\ref{purn}, iii).

In case of $\tau_1=\tau_2=\ldots=\tau_N=\tau$, the purity is given by
${\cal P}_{N,N}=\binom{4N}{2N}/\binom{2N}{N}^2/\cal{D}$  (Fig.\ref{purn}, iv). This can be verified by inserting
$\forall \tau_i=\tau$ in ${\cal P}_{N,N}$(Eq. (\ref{purint})) then taking the limit $\tau\rightarrow \infty$,
\begin{align}
{\cal P}_{N,N}&=\frac{1}{\cal D}\lim_{\tau\rightarrow\infty}\frac{\av{\cos^{4N}\frac{h\tau}{2}}}
{\av{\cos^{2N}\frac{h\tau}{2}}^2}\\ &=\frac{1}{\cal D} \frac{\sum_{s=0}^{4N}\binom{4N}{s}\av{e^{ih(s-2N)\tau}}}
{\bigl(\sum_{s=0}^{2N}\binom{2N}{s}\av{e^{ih(s-N)\tau}}\bigr)^2 }
=\frac{1}{\cal D}\frac{\binom{4N}{2N}}{(\binom{2N}{N})^2},
\label{eqtimes}
\end{align}
where in the last line
$\lim_{\tau\rightarrow\infty}\av{\exp{(ih n\tau)}}=\lim_{\tau\rightarrow\infty}
 \exp{(-n^2 \sigma^2 \tau^2/2)}=
\delta_{n,0}$
from (\ref{ident}) is employed.

For the case $\tau_1=2\tau_2=2^2\tau_3=\ldots=2^{N-1}\tau_N=\tau$, the asymptotic value of the purity
in the limit of $\sigma \tau \rightarrow \infty$ can be evaluated more systematically from the expression in (76)
rather than (82). The density matrix after the $N$ times measurements is given by
\begin{equation}
\rho_{N, N}=\frac{\cal N}{4^N} \sum_n p_n \hat{\rho}_n \frac{\sin^2 h_n \tau}{\sin^2 (h_n \tau/2^N)} \;,
\end{equation}
where the normalization constant ${\cal N}$ is determined by
\begin{equation}
1={\rm Tr} \; \rho_{N, N}= \frac{\cal N}{4^N} \sum_n p_n \frac{\sin^2 h_n \tau}{\sin^2 (h_n \tau/2^N)} \;.
\label{eq2}
\end{equation}
The last factor takes a large value about $4^N$ near $h_n \sim h_s=2^N s \pi/\tau (s \in \mathbb{Z})$ and can be
approximated as
\begin{equation}
\frac{\sin^2 h_n \tau}{\sin^2 (h_n \tau/2^N)} \cong 4^N \; \sum_{s \in \mathbb{Z}}
\frac{\sin^2 (h_n-h_s) \tau}{(h_n-h_s)^2 \tau^2} \;.
\end{equation}
In the limit of $\sigma \tau \rightarrow \infty$, the Gaussian
distribution $p_n$ is much broader than the last factor in (\ref{eq2}) and we have
\begin{align}
&1 \cong {\cal N} \sum_{s \in \mathbb{Z}}\frac{1}{\sqrt{2\pi} \sigma}
\exp\left[- \frac{h_s^2}{2\sigma^2}\right] \int_{-\infty}^{\infty} dh \;
\frac{\sin^2 (h-h_s) \tau}{(h-h_s)^2 \tau^2} \\
& \cong {\cal N} \sum_{s \in \mathbb{Z}}\frac{1}{\sqrt{2\pi} \sigma \tau}
\exp\left[- \frac{h_s^2}{2\sigma^2}\right] \int_{-\infty}^{\infty} dx \frac{\sin^2 x}{x^2} \\
& = {\cal N}  \sqrt{\frac{\pi}{2}}\frac{1}{\sigma \tau} \sum_{s \in \mathbb{Z}}
\exp\left[- \frac{h_s^2}{2\sigma^2}\right] \;.
\end{align}
Assuming furthermore $\sigma \tau \gg 2^N$, the sum over the integer $s$ can be replaced by an integral and
${\cal N}$ can be fixed as
\begin{align}
&1 \cong {\cal N} \sqrt{\frac{\pi}{2}} \frac{1}{\sigma \tau} \int_{-\infty}^{\infty} ds \;
\exp\left[- \frac{h_s^2}{2\sigma^2}\right] =\frac{{\cal N}}{2^N} \\
& \rightarrow {\cal N}= 2^N \;.
\end{align}
This result is equal to the exact result ${\cal N}=2^N$, i.e.,
\begin{align}
&1={\cal N} \sum_n p_n \cos^2 (\frac{h_n \tau_1}{2}) \cos^2 (\frac{h_n \tau_2}{2}) \cdots \cos^2
(\frac{h_n \tau_N}{2}) \\
&=\frac{{\cal N}}{4^N} \sum_{s_1=0}^{2} \sum_{s_2=0}^{2} \cdots \sum_{s_N=0}^{2} \binom{2}{s_1} \binom{2}{s_2} \cdots
\binom{2}{s_N} \; \delta((s_1-1)\tau_1 +(s_2-1)\tau_2 + \cdots (s_N-1)\tau_N) \\
&= \frac{{\cal N}}{4^N} 2^N  \longrightarrow {\cal N}=2^N \;.
\end{align}

Now that the density matrix is determined, the purity is calculated as
\begin{equation}
{\cal P}_{N, N}={\rm Tr} \; \sigma^2_{N, N}= \frac{1}{{\cal D}} \frac{{\cal N}^2}{16^N}
\sum_n p_n \frac{\sin^4 h_n \tau}{\sin^4 (h_n \tau/2^N)} \;.
\end{equation}
By the same arguments as above, we can approximate the last factor as
\begin{equation}
\frac{\sin^4 h_n \tau}{\sin^4 (h_n \tau/2^N)} \cong 16^N
\sum_{s \in \mathbb{Z}} \frac{\sin^4 (h_n-h_s) \tau}{(h_n-h_s)^4 \tau^4}
\end{equation}
and under the condition $\sigma \tau \gg 1$ we have
\begin{align}
&\sum_n p_n \frac{\sin^4 h_n \tau}{\sin^4 (h_n \tau/2^N)} \cong 16^N \sum_n p_n \; \sum_{s \in \mathbb{Z}}\;
\frac{\sin^4 (h_n-h_s) \tau}{(h_n-h_s)^4 \tau^4} \\
&\cong 16^N \; \sum_{s \in \mathbb{Z}} \frac{1}{\sqrt{2\pi}\sigma} \exp\left[- \frac{h_s^2}{2\sigma^2}\right]
\int_{-\infty}^{\infty} dh \; \frac{\sin^4 (h-h_s) \tau}{(h-h_s)^4 \tau^4} \\
&=16^N \; \sum_{s \in \mathbb{Z}} \frac{1}{\sqrt{2\pi}\sigma} \exp\left[- \frac{h_s^2}{2\sigma^2}\right]
\frac{1}{\tau} \int_{-\infty}^{\infty} dx \; \frac{\sin^4 x}{x^4} \\
&=16^N \frac{\sqrt{2\pi}}{3} \frac{1}{\sigma \tau} \; \sum_{s \in \mathbb{Z}}
 \exp\left[- \frac{h_s^2}{2\sigma^2}\right] \;.
\end{align}
Assuming $\sigma \tau \gg 2^N$, the summation over $s$ is replaced by an integral and we obtain
\begin{align}
&\cong 16^N \frac{\sqrt{2\pi}}{3} \frac{1}{\sigma \tau} \;
\int_{-\infty}^{\infty} ds \; \exp\left[- \frac{h_s^2}{2\sigma^2}\right] = \frac{2}{3} 8^N \\
& \rightarrow {\cal P}_{N, N} \cong \frac{{\cal N}^2}{{\cal D}} \frac{1}{16^N}\;
\sum_n p_n \frac{\sin^4 h_n \tau}{\sin^4 (h_n \tau/2^N)}=\frac{1}{{\cal D}} \frac{2}{3} \; 2^N \;.
\end{align}
This expression reproduces very well the result in Fig.\ref{purn}, i.


\end{document}